\documentclass[10pt, conference]{IEEEtran}
\IEEEoverridecommandlockouts
\usepackage{subcaption}
\usepackage{cite}
\usepackage{amsmath,amssymb,amsfonts}
\usepackage{array}
\usepackage{multirow}
\usepackage{diagbox}
\usepackage{algorithm}
\usepackage[noend]{algorithmic}
\usepackage{graphicx}
\usepackage{textcomp}
\usepackage{xcolor}
\usepackage{amsthm}
\usepackage{braket}
\usepackage{qcircuit}
\usepackage[english]{babel}
\usepackage{booktabs}
\usepackage{multirow}

\usepackage{hyperref}
\usepackage{float}

\usepackage[font=small]{caption}

\newcommand{\mlqls}{ML-SABRE}

\def\BibTeX{{\rm B\kern-.05em{\sc i\kern-.025em b}\kern-.08em
    T\kern-.1667em\lower.7ex\hbox{E}\kern-.125emX}}
\begin{document}

\title{
A High-Performance Multilevel Framework for Quantum Layout Synthesis}

\author{
\normalsize Shuohao Ping$^1$, Naren Sathishkumar$^1$, Wan-Hsuan Lin$^1$, Hanyu Wang$^1$, Jason Cong$^1$\\
\normalsize $^1$Computer Science Department, University of California, Los Angeles, CA, USA,\\ 
\normalsize \{sp1831,nks676,wanhsuanlin,hanyuwang\}@ucla.edu;cong@cs.ucla.edu
\vspace{-3mm}
}
\maketitle

\begin{abstract}
Quantum Layout Synthesis (QLS) is a critical compilation stage that adapts quantum circuits to hardware constraints with an objective of minimizing the SWAP overhead.
While heuristic tools demonstrate good efficiency, they often produce suboptimal solutions, and exact methods suffer from limited scalability.
In this work, we propose \mlqls, a high-performance multilevel framework for QLS that improves both solution quality and compilation time through a hierarchical optimization approach.
We employ the state-of-the-art heuristic method, LightSABRE, at all levels to ensure both efficiency and performance. 
Our evaluation on real benchmarks and hardware architectures shows that \mlqls\ decreases SWAP count by over 60\%, circuit depth by 17\%, and delivers a 60\% compilation time reduction compared to state-of-the-art solvers. 
Further optimality studies reveal that \mlqls\ can significantly reduce the optimality gap by up to 82\% for SWAP count and 49\% for circuit depth, making it well-suited for emerging quantum devices with increasing size and architectural complexity.
\mlqls\ is open source at \href{https://github.com/UCLA-VAST/multilevel-sabre}{https://github.com/UCLA-VAST/multilevel-sabre}.

\end{abstract}

\section{Introduction}

\begin{figure*}[htbp]
\centering
\includegraphics[width=\textwidth]{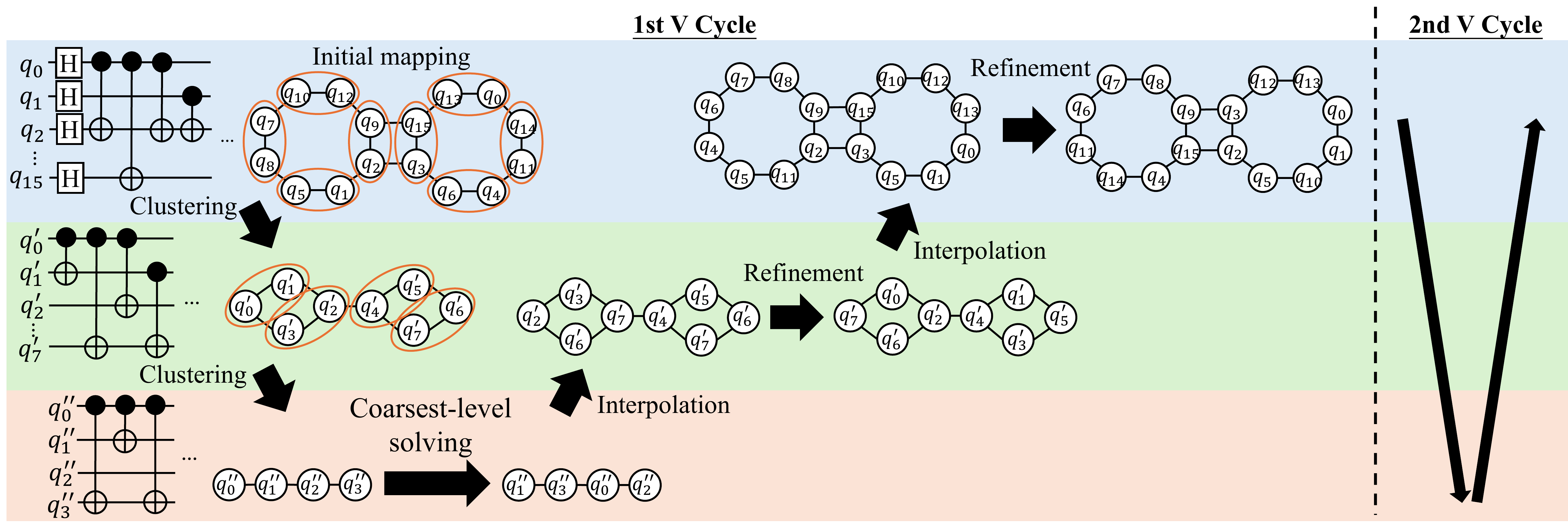}

\caption{Overview of multilevel framework for QLS.}
\label{fig:overview}
\end{figure*}

Due to the mature fabrication technology and high clock rate, superconducting qubits have emerged as one of the promising solutions to realize quantum computing  
Their hardware architecture is typically represented by a coupling graph, where vertices denote physical qubits and edges indicate pairs that can directly perform two-qubit gates. 
Because of frequency collisions and crosstalk, these coupling graphs are often sparse, limiting direct gate execution \cite{rigetti,chow2021ibmeagle,google2024willow}. 
However, quantum circuits are typically generated assuming full connectivity, i.e., any two qubits may interact. 
This mismatch between circuit and hardware connectivity necessitates quantum layout synthesis (QLS), which is the process to accommodate circuit connectivity to the device connectivity.
The task for QLS is to find the mapping from program qubits in the given circuit to the target coupling graph. 
Since a single static mapping often cannot support all required two-qubit interactions, QLS dynamically updates the mapping by inserting logical SWAP gates. 
Nevertheless, these SWAP gates negatively impact circuit performance in two key ways. 
First, they introduce additional operations, increasing the chance of errors as a result of imperfect gate fidelity. 
Second, SWAPs on the critical path extend the circuit execution time, worsening the decoherence effects.
Thus, the main objective of the QLS stage is to minimize the number of inserted SWAPs.

The QLS problem has been proved to be NP-hard~\cite{siraichi_qubitallocation_2018}, 
and the current approaches can be broadly categorized into exact and heuristic methods. 
Exact approaches model QLS as a constraint satisfaction problem to find optimal solutions, but they struggle to scale beyond small- to medium-sized circuits~\cite{wille2014optimal,bhattacharjee2019muqut,wille2019mapping,tan2020olsq, tan2021olsqga, zhang_time-optimal_2021, molavi2022satmap,nannicini2022optimal, lin2023olsq2}.
Despite their high solution quality, exact methods become impractical in the face of rapidly advancing hardware with hundreds of qubits, as they often incur compilation times on the order of hours. This long overhead makes them unsuitable for large-scale or time-sensitive applications.
To handle larger instances, heuristic algorithms have been introduced, offering significantly faster runtime~\cite{siraichi_qubitallocation_2018,ho2018cirq,zulehner2018mapping_to_ibm_qx,web18-ibm-qiskit, zulehner_efficient_2019, siraichi_qubit_2019, li_sabre_2019,murali_formal_2019,sivarajah_tket_2020,kole_improved_2020,liu_notallswaparethesame_2022,wu_robust_2022,fan_QLSML_2022,huang2022reinforcement,park2022fsqm,huang2024ctqr,huang2024dear,lin2024mlqls}.
However, their scalability comes at the potential cost of solution quality.
Recent studies have shown that heuristic tools can exhibit large optimality gaps on both SWAP-free instances~\cite{tan2020queko} and those requiring SWAP insertion~\cite{qubikos}, highlighting limitations in their solution quality despite fast runtimes. 
Specifically,\cite{qubikos} reports an optimality gap as high as 234$\times$ in SWAP count, while\cite{tan2020queko} shows a gap of up to 6$\times$ in circuit depth for large circuit and device.

To address the scalability-quality trade-off, recent work, ML-QLS, has proposed a multilevel framework for solving QLS using a hybrid approach that combines exact and heuristic methods~\cite{lin2024mlqls}. 
Multilevel techniques demonstrate a strong performance in large-scale optimization across domains such as VLSI design, where they are used for circuit partitioning~\cite{alpert1997multilevel,cong2004edge}, 
placement~\cite{chan_enhanced_2003,chen2005ntuplace,chan2005multilevel,cheng2018replace,leong2009replace,chan2000multilevel,chan2005mpl6}, 
and routing~\cite{karypis1997multilevel,cong2005thermal,ou2012non,lin2002novel,liu2020cugr} to manage designs with millions of transistors. 
In ML-QLS, clustering is utilized to reduce the problem size until the coarsest level is tractable for an exact solver. 
From there, the solution is progressively refined at each finer level using a simulated annealing-based initial mapping strategy and an A* search-based SWAP insertion algorithm. 
Figure~\ref{fig:overview} illustrates an example of this multilevel solving process. 
To ensure the exact solver runs within a reasonable time, clustering continues until the circuit is trimmed below a certain threshold. However, this aggressive reduction can lead to over-clustering, potentially discarding important structural information from the original circuit and coupling graph. 
Additionally, while ML-QLS achieves SWAP count improvements in some cases, its overall efficiency remains constrained, particularly due to the A* search component suffering from state explosion as the problem size grows.
For example, ML-QLS takes 30 minutes to solve a Quantum Fourier Transform (QFT) circuit featuring 29 qubits and 406 two-qubit gates,
compared to commercial tools such as Qiskit  LightSABRE~\cite{zou2024lightsabre} (referred as SABRE in the rest of the paper), which typically finish in one second. 

To this end, we propose \mlqls, a multilevel-based QLS tool that significantly reduces the optimality gap while achieving shorter runtime by leveraging the state-of-the-art heuristic tool, SABRE, as the underlying QLS engine.
Improving upon ML-QLS, our approach introduces the following key enhancements:
\begin{itemize}
    \item A topology-aware initial embedding strategy that aligns circuit and device structures, significantly enhancing clustering quality.
    \item An advanced multilevel flow incorporating multiple V-cycles to iteratively improve clustering and a restart mechanism to escape local minimum.
    \item An adaptive clustering strategy to prevent overclustering, which may undermine the effectiveness of the coarser-level solving.
    \item A flexible refinement scheme that explores multiple interpolation outcomes, further enhanced by random trial-based exploration to effectively traverse the solution space.
\end{itemize}
Due to versatility of the multilevel framework in QLS, as heuristic algorithms improve, they can be seamlessly integrated as the underlying engine to further enhance performance.


\section{Background}
\label{sec:background}

\subsection{Quantum Layout Synthesis}

\begin{figure}[htbp]
\centering

\begin{subfigure}[b]{0.55\linewidth}
    \centering
    \includegraphics[width=\linewidth]{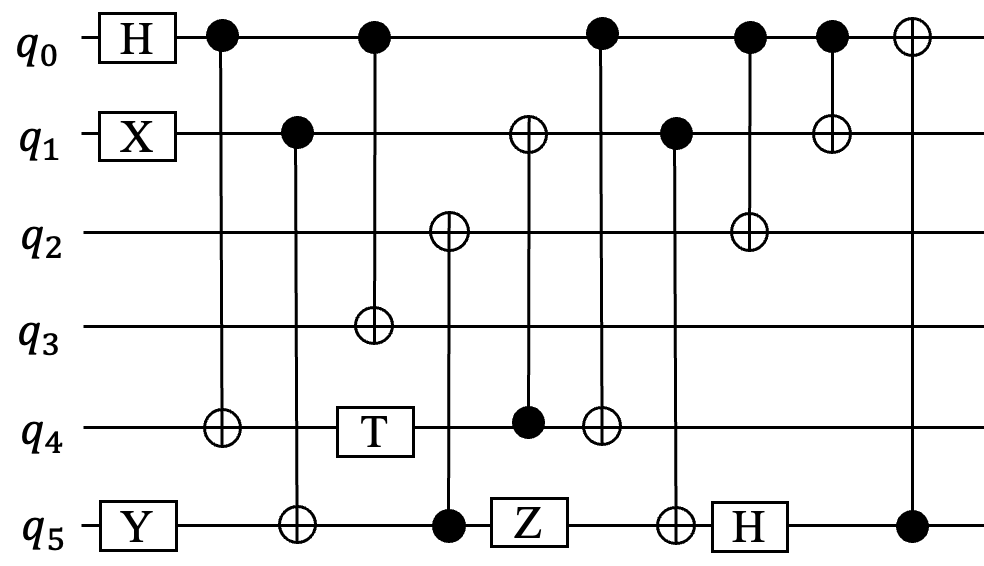}
    \caption{}
    \label{fig:circuitexample}
\end{subfigure}
\hfill
\begin{subfigure}[b]{0.43\linewidth}
    \centering
    \includegraphics[width=0.8\linewidth]{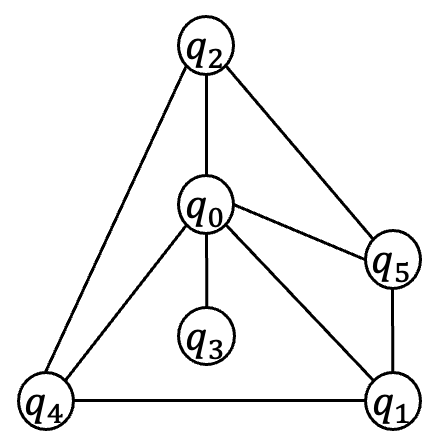}
    \caption{}
    \label{fig:interactionexample}
\end{subfigure}



\begin{subfigure}[b]{0.35\linewidth}
    \centering
    \includegraphics[width=\linewidth]{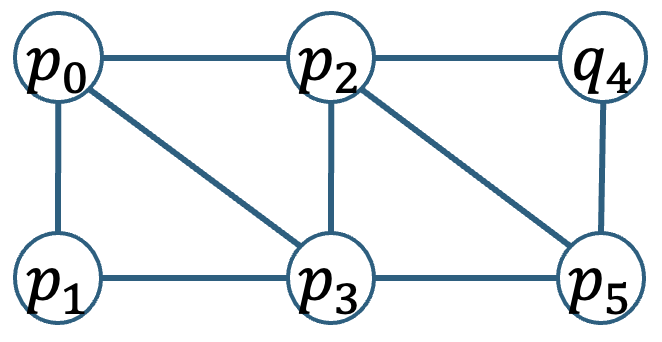}
    \caption{}
    \label{fig:device example}
\end{subfigure}
\hfill
\begin{subfigure}[b]{0.63\linewidth}
    \centering
    \includegraphics[width=\linewidth]{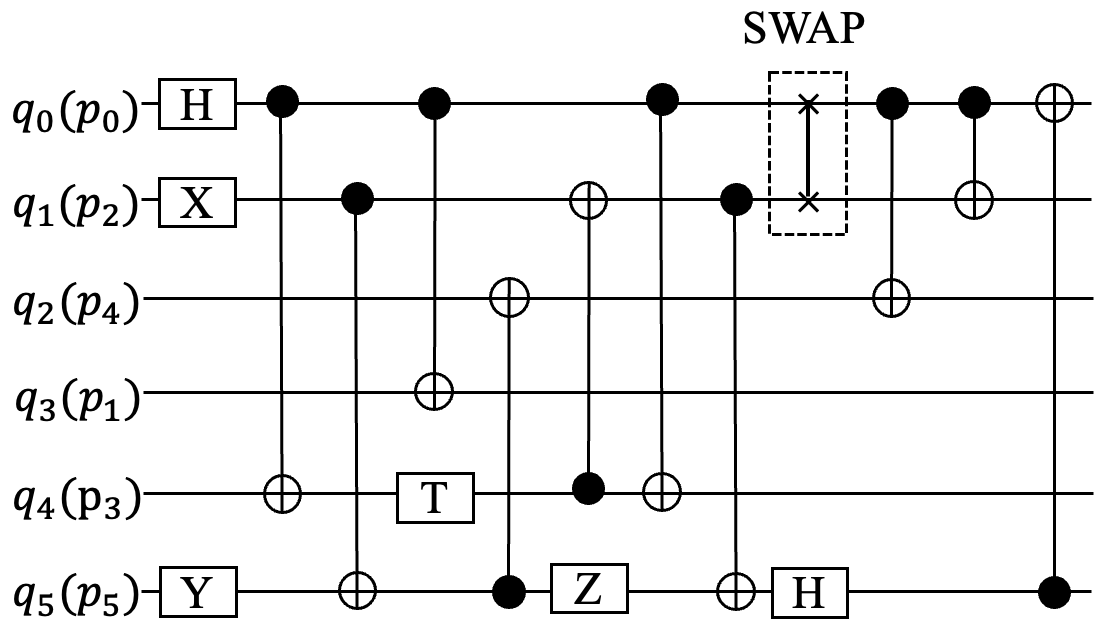}
    \caption{}
    \label{fig:compiled}
\end{subfigure}

\caption{A QLS example. (a) A quantum circuit. (b) The interaction graph for the circuit.
(c) A device coupling graph.
(d) Compiled circuit with one SWAP gate and initial mapping.}
\label{fig:fullexample}
\end{figure}

The inputs to QLS for superconducting qubits are a quantum circuit $C$ and a coupling graph $G_C$.
Figure~\ref{fig:circuitexample} demonstrates a six-qubit quantum circuit with six single-qubit gates and ten two-qubit gates, 
and Figure~\ref{fig:device example} shows a 6-qubit couping graph.
In this paper, we denote the set of physical qubits in the coupling graph by $P$  and the set of edges by $E$.
For a circuit, we denote the set of program qubit by $Q$, and the set of gates by $\mathcal{G}=\mathcal{G}_1 \cup \mathcal{G}_2$, 
where $\mathcal{G}_1$ and $\mathcal{G}_2$ are the set of single-qubit and two-qubit gates, respectively.
Based on $\mathcal{G}_2$, we can define the circuit interaction graph $G_I$, where each program qubit is a vertex, and two qubits share an edge if there is a two-qubit gate acting on them, as seen in Figure~\ref{fig:interactionexample}.

The output of the QLS problem is a compiled circuit with inserted SWAP gates, and an initial mapping from program qubits to physical qubits $f:Q\rightarrow P$. 
Figure~\ref{fig:compiled} depicts a QLS result using one SWAP to map the circuit in Figure \ref{fig:circuitexample} to the coupling graph in Figure \ref{fig:device example}. 
The result is valid if 1) the mapping is injective, and 2) the gate execution order is preserved, and 3) two-qubit gates happen on adjacent physical qubits.




\subsection{Multilevel Framework}
The multilevel framework is effective for large-scale optimization problems, where solving the finest-level instance directly is intractable and heuristics risk getting trapped in local minima.
The process begins with \emph{clustering} to create a hierarchy of coarser approximations until the coarsest-level problem is small enough to solve optimally. 
The \emph{coarsest-level solving} is expected to generate an optimal solution.
The long runtime overhead by an exact solver is justified by the high-quality global optimization information, which provides valuable guidance for subsequent levels.
Based on the coarsest-level solution, we first perform \emph{interpolation} to project the solution to the finer solution space. 
Then, guided by a good coarser-level solution, \emph{refinement} often requires only local optimization and converges fast.
Moreover, refinement is typically efficient and repeated at each level as we move back to the finest problem. 
This progression from large to small and back to large problem sizes forms the characteristic ``V'' shape of the \emph{V cycle}.

Figure~\ref{fig:overview} illustrates an example solving process of multilevel framework for QLS based on~\cite{lin2024mlqls}.
Clustering is applied to both the circuit and the coupling graph to reduce the problem size.
Note that a circuit-informed device clustering strategy is proposed to increase similarity between the coupling graph and the qubit interaction graph.
Based on a given mapping from program qubits to physical qubits, we first cluster program qubits that are mapped to the adjacent physical qubits, and then the corresponding physical qubits form a coarser physical qubit.
In addition, a compression rate of two is selected to ensure a zero-cost execution when performing intra-clustering operations for a coarser physical qubit.
After the problem size is decreased to around 10 qubits, a solver-based method is applied at the coarsest level.
Based on the mapping from the coarser level, an initial mapping is derived during interpolation, and further improved through refinement operations. 
After one V cycle, one can proceed to having additional V cycles to gradually improve the quality of the initial mapping, and hence have better clustering decisions.

\section{ML-SABRE}
 \label{sec:mlqls}

\subsection{Overview}
\label{sec:overview}

\begin{algorithm}[htbp]
\caption{\mlqls 
}
\label{alg:ml_framework}
\begin{algorithmic}[1]
\REQUIRE Circuit $C$, Coupling Graph $G_C$, Number of Cycle $d$
\ENSURE Compiled Circuit $C_r$

\STATE $f \leftarrow$ \textsc{InitialEmbedding}$(C, G_C)$ \label{line:embedding}
\STATE Initialize empty stack $\mathcal{S}$ and empty set $\mathcal{T}$
\STATE $\mathcal{S}.\text{push}(C, G_C, f)$
\FOR {cycle = 1: $d$}
\WHILE{not \textsc{Embeddable}$(C, G_C, f)$} \label{line:embedable}
    \STATE $(C, G_C, f) \leftarrow$ \textsc{Cluster}$(C, G_C, f)$ \label{line:clustering}
    \STATE $\mathcal{S}.\text{push}(C, G_C, f)$
\ENDWHILE

\STATE $\mathcal{S}.\text{pop}()$ \COMMENT{Discard top level (embeddable)} \label{line:pop1}
\STATE $(C, G_C, \_) \leftarrow \mathcal{S}.\text{pop}()$ \label{line:pop2}
\STATE $f, \_ \leftarrow$ \textsc{CoarsestLevelSolve}$(C, G_C)$ \label{line:cls} 

\WHILE{$\mathcal{S}$ is not empty}
    \STATE $(C, G_C, \_) \leftarrow \mathcal{S}.\text{pop}()$
    \STATE $f \leftarrow$ \textsc{Interpolation}$(C, G_C, f)$ \label{line:interpolation} 
    \STATE $f, C' \leftarrow$ \textsc{Refinement}$(C, G_C, f)$ \label{line:refinement} 
\ENDWHILE
\STATE Update $C_r$ by $C'$ if SWAP count reduces
\IF{$f$ in $\mathcal{T}$}
\label{line:converge_start}
\STATE Generate a random mapping $f \notin \mathcal{T}$ 
\label{line:converge_end}
\ENDIF
\label{line:next_cycle_start}
\STATE $\mathcal{S}.\text{push}(C,G_c,f)$
\STATE $\mathcal{T}.\text{add}(f)$
\label{line:next_cycle_end}
\ENDFOR
\RETURN $C_r$
\end{algorithmic}
\end{algorithm}

In this section, we present \mlqls, a scalable QLS 
built upon a multilevel framework. 
To ensure practical compilation time, we adopt heuristic methods for both coarsest-level solving and refinement across all levels. 
The use of heuristics at the coarsest level is supported by findings in~\cite{qubikos}, where the authors show that state-of-the-art heuristic methods, such as SABRE, achieve near-optimal results when the problem size is small.
The multilevel framework also employs an iterative V cycle to gradually enhance solution quality. 
By repeatedly coarsening and refining the problem, each cycle incrementally improves the mapping, allowing the algorithm to escape local minima and converge toward higher-quality solutions.
Algorithm~\ref{alg:ml_framework} shows our flow.

To generate a high-quality clustering for the first V cycle, we propose an effective embedding strategy to map the circuit to the device based on the circuit connectivity (line~\ref{line:embedding}).
The initial mapping is essential for guiding the clustering algorithm to group qubits with high interaction frequencies while aligning with both the circuit and device topology~\cite{lin2024mlqls}. 
This approach improves the effectiveness of the multilevel framework by improving the quality of the coarsening process.

Based on the initial mapping, an iterative clustering strategy is applied to generate coarser circuits and coupling graphs.
In each iteration, we construct a coarser circuit, a coupling graph, a trivial (identity) qubit mapping 
(line~\ref{line:clustering}). A trivial mapping is possible because we need to label each program and physical coarser after clustering, and we can always label them in a way that the mapping after clustering is trivial.
The clustering process terminates when the trivial mapping achieves a perfect embedding. 
That is, the circuit can be executed on the device under the current mapping without SWAP insertion (line~\ref{line:embedable}).
Then, we select the previous clustering stage as our coarsest level (line~\ref{line:pop1}-\ref{line:pop2}),
as solving the embeddable case does not yield meaningful insights for global optimization. 
This is because once the trivial mapping reaches optimal, coarsest-level solving will always return the same trivial mapping as output without exploring alternative mappings that could lead to better results.
In contrast, when the trivial mapping is imperfect, coarsest-level solving has the opportunity to search for other mapping, enabling a more global search.

At the coarsest level, we solve the mapping problem with a heuristic approach to obtain a near-optimal solution (line~\ref{line:cls}) and then iteratively project the solution to a finer level through interpolation (line~\ref{line:interpolation}). 
Next, refinement is employed to improve the solution (line~\ref{line:refinement}). 
We repeat interpolation and refinement until a valid solution to the finest level is reached,
which completes a single V cycle.
Starting the second V cycle, we use the qubit mapping from the first cycle as the initial mapping (line \ref{line:next_cycle_start}--\ref{line:next_cycle_end}). 
If the mapping from the current cycle already appears in previous cycles,
it indicates convergence of the multilevel optimization.
In such cases, we restart the optimization with a new initial mapping (Line~\ref{line:converge_start}--\ref{line:converge_end}).
The subsequent sections offer a detailed discussion on each step in the multilevel framework.
Section~\ref{subsec:initial_embedding}  presents our initial embedding strategy. The clustering algorithm is detailed in Section~\ref{subsec:clustering}, followed by the interpolation method in Section~\ref{subsec:interpolation}. In Section~\ref{subsec:refinement}, we describe the refinement process. Finally, we conclude the section with a time complexity analysis of \mlqls\ in Section~\ref{subsec:runtime}.

\subsection{Initial Embedding}
\label{subsec:initial_embedding}


As discussed in Section~\ref{sec:overview}, clustering aims to align the circuit and device topologies. 
Achieving this alignment requires a good initial embedding.
An effective initial embedding should map as many edges from the interaction graph onto the physical device as possible, thereby improving the topological similarity before co-clustering.
When the circuit interaction graph can be directly embedded into the hardware coupling graph, the problem becomes a subgraph isomorphism problem, solvable by algorithms such as VF2~\cite{VF2} with exponential time complexity. 
However, when the graphs are not isomorphic, algorithms like VF2 fail to solve the problems in a reasonable time. 
To address this, we propose two heuristic algorithms that generate high-quality initial mappings in polynomial time that target specific circuit connectivity, i.e., line and star-like structures. 
These heuristics aim to preserve the structure of the circuit interaction graph, even when it cannot be perfectly embedded into the device topology.

\begin{figure}
\centering

\begin{subfigure}[t]{0.48\linewidth}
    \centering
    \raisebox{3em}{\includegraphics[width=\linewidth]{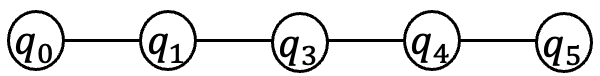}}
    \caption{}
    \label{fig:linestructure}
\end{subfigure}
\hfill
\begin{subfigure}[t]{0.48\linewidth}
    \centering
    \includegraphics[width=0.7\linewidth]{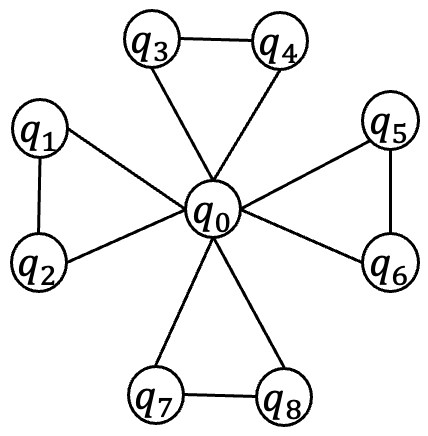}
    \caption{}
    \label{fig:starlikestrcture}
\end{subfigure}

\caption{Example of circuit interaction graphs (a) A 6-qubit Ising model.
(b) A 9-qubit SWAP test.}
\label{fig:strctures}
\end{figure}

The line structure features two-qubit interaction on only linear nearest neighbors,
e.g., the Ising model and state preparation circuits.
Figure~\ref{fig:linestructure} depicts an example interaction graph.
The star-like circuits characterize by a central qubit that acts as a hub of global entanglement, interacting with all other qubits. 
The remaining qubits typically engage in more localized interactions, forming a star-shaped interaction graph, as illustrated in Figure~\ref{fig:starlikestrcture}. 
Example circuits include the Bernstein–Vazirani algorithm~\cite{bv} and the SWAP test.

If a circuit contains these structures as substructures, we can isolate them and map each substructure to the coupling graph utilizing our heuristic. 
For circuits without these specific structures, we apply a general strategy by mapping all program qubits to a densely connected region, optimizing the likelihood of satisfying the circuit's topology.

\begin{figure}[ht]
\centering

\begin{subfigure}[b]{0.48\linewidth}
    \centering
    \includegraphics[width=0.9\linewidth]{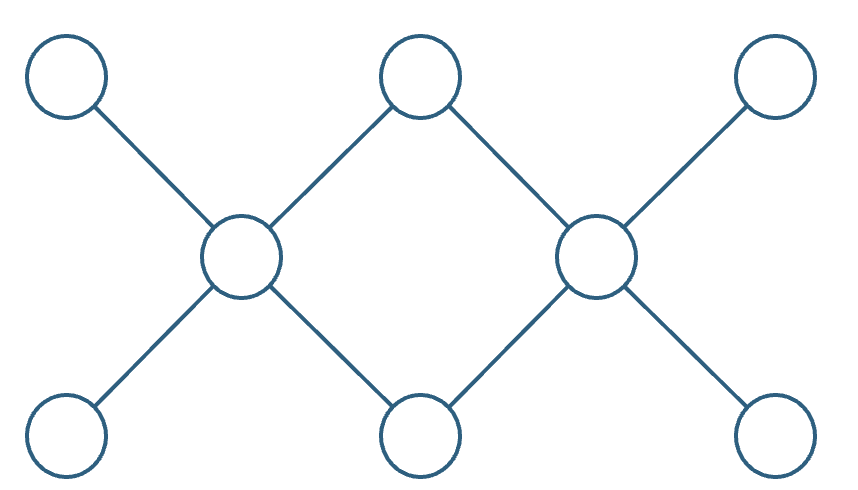}
    \caption{}
    \label{fig:linea}
\end{subfigure}
\hfill
\begin{subfigure}[b]{0.48\linewidth}
    \centering
    \includegraphics[width=0.9\linewidth]{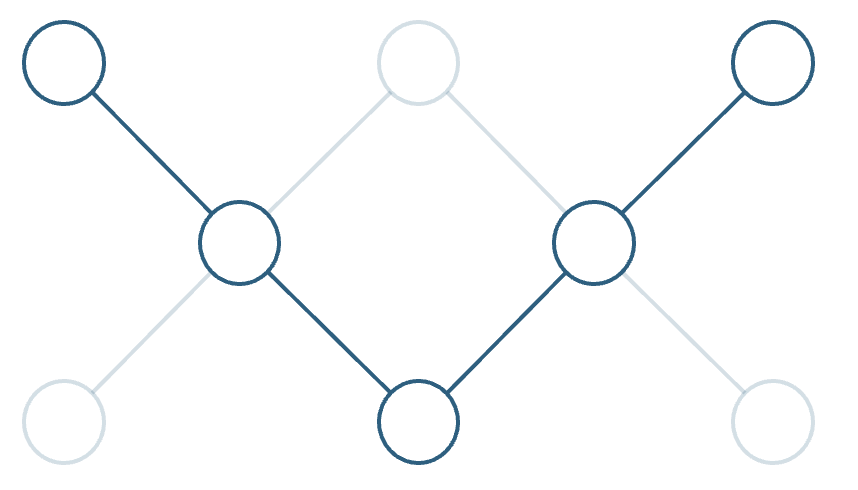}
    \caption{}
    \label{fig:lineb}
\end{subfigure}


\begin{subfigure}[b]{0.48\linewidth}
    \centering
    \includegraphics[width=0.9\linewidth]{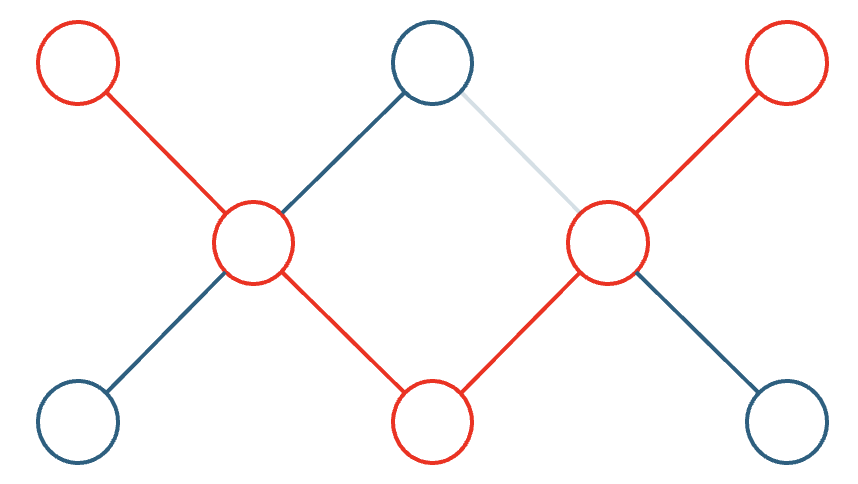}
    \caption{}
    \label{fig:linec}
\end{subfigure}
\hfill
\begin{subfigure}[b]{0.48\linewidth}
    \centering
    \includegraphics[width=0.9\linewidth]{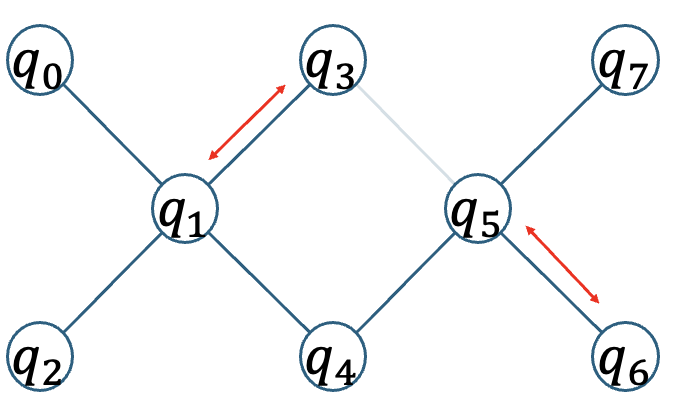}
    \caption{}
    \label{fig:lined}
\end{subfigure}

\caption{Example of embedding for a 8-qubit circuit with linear connectivity. 
(a) A coupling graph. 
(b) A longest path in the coupling graph. 
(c) A tree
 created by extending the longest path with neighboring qubits. 
(d) Initial embedding. 
The red arrows indicate the required SWAP operations.
}
\label{fig:lineembeding}
\end{figure}

\begin{figure}[ht]
\centering

\begin{subfigure}[b]{0.48\linewidth}
    \centering
    \includegraphics[width=0.9\linewidth]{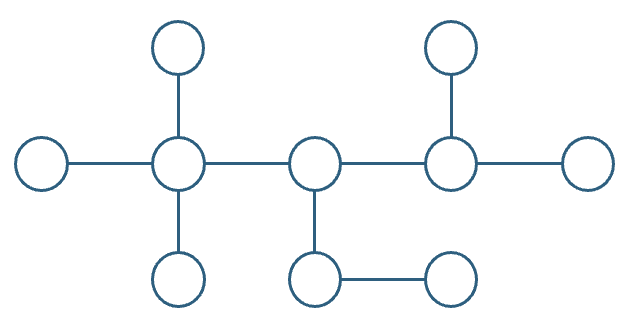}
    \caption{}
    \label{fig:stara}
\end{subfigure}
\hfill
\begin{subfigure}[b]{0.48\linewidth}
    \centering
    \includegraphics[width=0.9\linewidth]{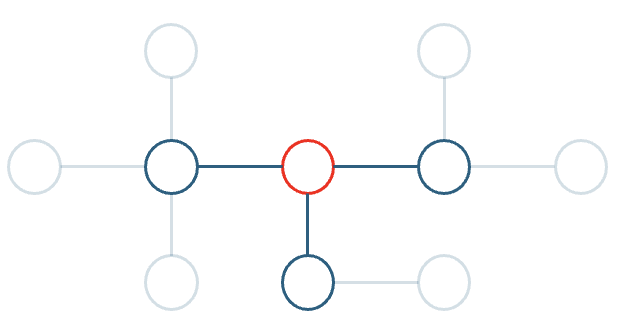}
    \caption{}
    \label{fig:starb}
\end{subfigure}


\begin{subfigure}[b]{0.48\linewidth}
    \centering
    \includegraphics[width=0.9\linewidth]{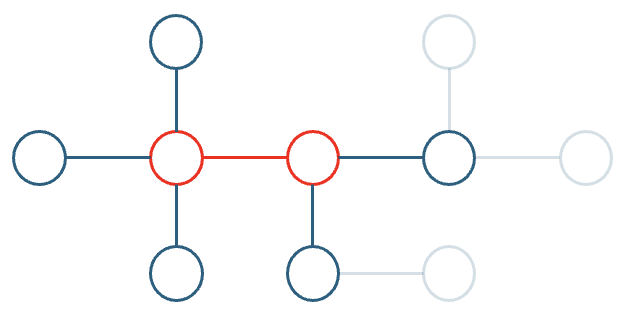}
    \caption{}
    \label{fig:starc}
\end{subfigure}
\hfill
\begin{subfigure}[b]{0.48\linewidth}
    \centering
    \includegraphics[width=0.9\linewidth]{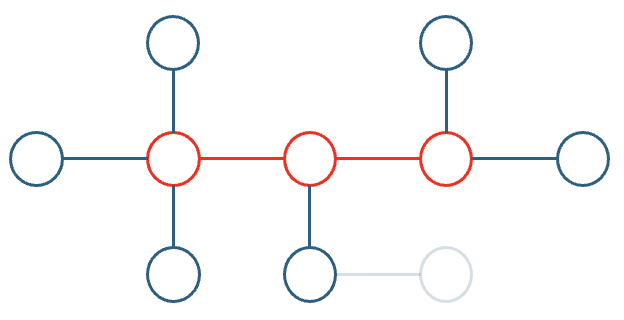}
    \caption{}
    \label{fig:stard}
\end{subfigure}

\begin{subfigure}[b]{0.48\linewidth}
    \centering
    \includegraphics[width=0.9\linewidth]{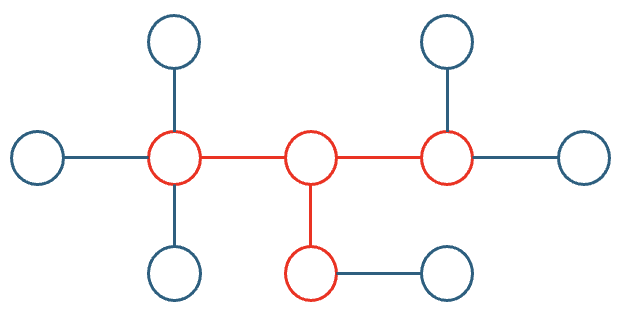}
    \caption{}
    \label{fig:stare}
\end{subfigure}
\hfill
\begin{subfigure}[b]{0.48\linewidth}
    \centering
    \includegraphics[width=0.9\linewidth]{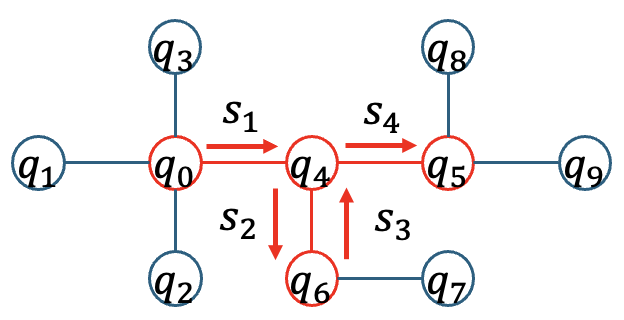}
    \caption{}
    \label{fig:starg}
\end{subfigure}
\caption{Example of embedding for a 10-qubit circuit with star-like connectivity, where $q_0$ interacts with $q_1$ to $q_9$.
(a) A coupling graph. 
(b) The starting point of the walk (in red) with its neighbors highlighted in blue.
(c) The first expansion step. 
The walk is extended leftward to include more qubits. 
(d) The second expansion step, where the walk is expanded rightward. 
(e) Select a new qubit in the extended set as the starting point. 
(f) Initial embedding. 
Red nodes indicate the walk of $q_0$, 
and red arrows represent the  SWAP operations executed in the order $s_1$ to $s_4$.}
\label{fig:starembeding}
\end{figure}

\subsubsection{Embedding for Circuit with Line Structure}
 \label{subsubsec:line mebedding}
To preserve the circuit's linear topology, we use a greedy heuristic to find the longest path in the coupling graph.
Specifically, we run DFS from every node to extract the longest path, and select the longest as the physical mapping path.
Figure~\ref{fig:linea} shows a coupling graph, and Figure~\ref{fig:lineb} exhibits the longest path selected by the above method.

If the path length meets or exceeds the number of program qubits, we directly map them onto the path, yielding a SWAP-free initial embedding.
Otherwise, we augment the path employing BFS from its nodes to include nearby qubits, appending them after their neighbors to form a locally linear tree structure.
Figure~\ref{fig:linec} shows an expansion example with the original longest path marked in red. 
Once enough physical qubits are gathered, we map program qubits to the deformed path following their order in the interaction graph, prioritizing mapping to side branches before the main path to preserve local linearity.
Figure~\ref{fig:lined} gives an example of mapping a line-structure circuit with 8 qubits and two-qubit gates acting on $q_i$ and $q_{i+1}$, to the device coupling graph.
Considering a coupling graph with $n$ physical qubits and maximum degree bounded by some constant, the complexity for this algorithm is $O(n^2)$: each DFS takes $O(n)$ time and is repeated for $n$ physical qubit.


\subsubsection{Embedding for Circuit With Star-like Structure}
\label{subsubsec:star mebedding}
For star-like circuits, we aim to place the center qubit where it can efficiently interact with all leaf qubits by traversing a walk on the coupling graph and interacting with neighboring qubits. 
Leaf qubits are then mapped to the ``extended set'', which are the qubits on the walk or adjacent to it, so the center qubit can interact with them as it traverses the walk.

Our algorithm grows a path originated from a qubit on the coupling graph by extending it at either end with neighboring nodes that maximizes the size of the extended set. 
The path expansion terminates when the size of the extended set cannot be increased.
For instance, 
Figure~\ref{fig:starembeding} illustrates an example of embedding a star-like circuit, where the center qubit $q_0$ interacts with eight leaf qubits $q_1$ to $q_9$. 
The two-qubit gates are executed sequentially in the order $(q_0,q_1), (q_0,q_2), \ldots, (q_0,q_9)$, 
and Figure~\ref{fig:stara} specifies the coupling graph.
The expansion starts from the red node (starting point), with its neighbors marked in blue (Figure~\ref{fig:starb}).
Next, the path is extended by adding the left-side qubits, as this choice increases the extended set by three, compared to only two when expanding to the right (Figure~\ref{fig:starc}). Figure~\ref{fig:stard} shows another expansion, and after this step, neither end of the path has neighbors that can enlarge the extended set,
marking the end of the expansion process.
If the extended set does not contain sufficient physical qubits, 
we select a qubit from the extended set but not on the  path to restart the expansion, as seen in Figure~\ref{fig:stare}. 
The process terminates if the extended set contains enough physical qubits for the circuit.
We greedily select every node in the coupling as the starting point, and only keep the walk with minimal length.

When embedding program qubits to physical qubits, we sort the leaf qubits based on their first appearance in the quantum circuit.
The center qubit is assigned to the first node on the walk. Then, for each node along the walk, we allocate its unassigned neighbors to the sorted leaf qubits. This is to minimize the chance of moving the center qubit back and forth on the walk when it interacts with the leaf qubit as the circuit progress.
Figure~\ref{fig:starg} exhibits an example embedding.
The worst-case complexity for this algorithm is $O(n^3)$ for a coupling graph with $n$ physical qubits when a single path cannot accommodate all program qubits, necessitating multiple starting points to be evaluated.
With $O(n)$ starting points (to restart expansion), because each requires $O(n)$ time to evaluate the extended set, and the expansion process repeats up to $n$ times (for greedy selection), the total worst-case complexity is $O(n^3)$.

\subsection{Clustering}
\label{subsec:clustering}
With a good initial embedding, we employ clustering to generate a problem hierarchy. 
The goal is to compress both the quantum circuit and the coupling graph while retaining the essential structural properties required for effective qubit mapping.
Specifically, we group strongly interacting program qubits into the same cluster as qubits in a grouped cluster that remains physically close, thereby reducing the need for SWAPs. 
By clustering them early in the multilevel process, we increase the likelihood that they will remain near each other in the final mapping, which minimizes routing overhead and improves overall mapping quality.

On the hardware side, clustering is restricted to connected physical qubits, ensuring that program clusters map to well-connected physical regions and reduce intra-cluster routing costs. 
To satisfy logical interaction locality and hardware connectivity, we introduce a circuit-device co-clustering strategy guided by an initial mapping. 

We first assign a weight to each edge in the coupling graph based on the interaction frequency between the mapped program qubit pair, reflecting how circuit interactions are distributed over the hardware.
To group frequently interacting qubit pairs into the same cluster and prevent unbalanced clusters that could degrade the effectiveness of coarsening, we perform a maximum-cardinality matching with maximum weight on this graph.
The matched physical qubits and the corresponding program qubits form a coarser physical and program qubit cluster, respectively.
Then, we construct the coarse-level circuit as follows: if a two-qubit gate in the finer-level circuit acts on different clusters, a corresponding gate is added between the corresponding coarser-level qubits. 
Otherwise, it is omitted.
The coarse coupling graph is constructed in a similar fashion. For any pair of coarse physical qubits, if there exists an edge in the fine-level coupling graph connecting nodes from each of the two corresponding clusters, then an edge is added between the coarse qubits in the coarse-level coupling graph.

\begin{figure}[ht]
\centering

\begin{subfigure}[t]{0.3\linewidth}
    \centering
    \includegraphics[width=\linewidth]{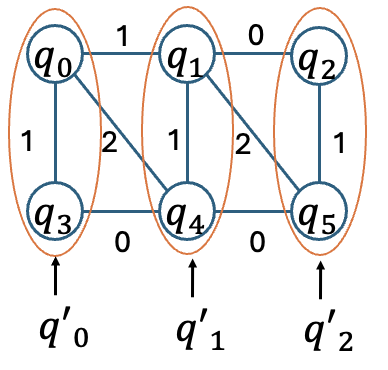}
    \caption{}
    \label{fig:cluster2}
\end{subfigure}
\hfill
\begin{subfigure}[t]{0.3\linewidth}
    \centering
    \raisebox{1em}{\includegraphics[width=\linewidth]{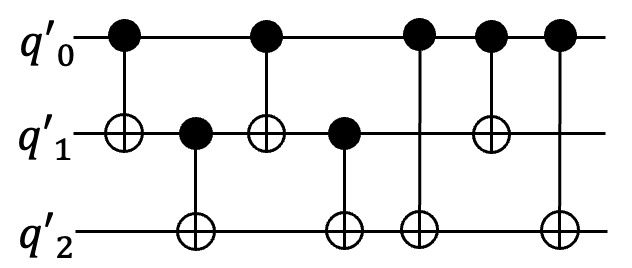}}
    \caption{}
    \label{fig:cluster3}
\end{subfigure}
\hfill
\begin{subfigure}[t]{0.3\linewidth}
    \centering
    \raisebox{1.5em}{\includegraphics[width=0.9\linewidth]{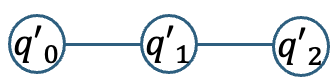}}
    \caption{}
    \label{fig:cluster4}
\end{subfigure}

\caption{
(a) The device coupling graph shown in Figure.\ref{fig:device example} with an initial mapping. The number on each edge is the weight derived from Figure.\ref{fig:circuitexample}. The orange circle denotes the matching result, where qubits within the same circle will be clustered into the same coarser qubit. (b). The quantum circuit after clustering. (c). The device coupling graph and corresponding mapping after clustering.}
\label{fig:cluster}
\end{figure}

Figure~\ref{fig:cluster} illustrates the construction of the coarser-level circuit and coupling graph using the circuit in Figure~\ref{fig:circuitexample} and the device in Figure~\ref{fig:device example}. 
In Figure~\ref{fig:cluster2}, the coupling graph is shown with an initial mapping and edge weights derived from the circuit.
The orange circles indicate the maximum weight and cardinality matching, 
and qubits in one circle form a coarser qubit. 
Note that if we perform maximum weight matching without considering cardinality, the resulting clusters would be $\{q_3\}, \{q_0,q_4\},\{q_1,q_5\},\{q_2\}$, which is an unbalanced configuration consisting of clusters with size two or one.
This imbalance increases the likelihood of mapping a coarse program qubit
to a coarse physical qubit with an insufficient number of physical qubits, which should be prevent as it may lead to invalid mapping in the finer level.
Finally, Figure~\ref{fig:cluster3} and \ref{fig:cluster4} displays the coarser circuit and coupling graph.

In~\cite{lin2024mlqls}, the author proposed to iteratively cluster the problem until the instance is small enough to be solved by an SMT solver. 
However, this approach risks over-clustering, where the coarsest-level mapping becomes trivial, providing no meaningful guidance for refining the finer-level mapping.
For instance, if we apply another round of clustering, the resulting coupling graph would consist of two nodes connected by one edge. 
At this level, any mapping would satisfy the circuit connectivity constraints, yielding no valuable insights for extracting global information.
In this case, Figure~\ref{fig:cluster4} illustrates the smallest instance that still requires SWAP gates, making it the coarsest level at which optimized mapping can provide meaningful guidance.

To address this over-clustering problem, we offer another strategy for iterative clustering. After each clustering step, we check whether the trivial mapping, obtained from the co-clustering stage, satisfies all the connectivity constraints imposed by the clustered circuit. If it does, this indicates that the problem has been reduced too far, and we discard this level. Instead, we use the previous level, where trivial mapping fails to meet the constraints, as the final coarse level for refinement.

For a  circuit with $g$ gates and coupling graph with size $n$, each clustering step requires $O(g)$ time to assign edge weights and $O(n^3)$ time to solve maximum cardinality and weight matching matching
with the Blossom algorithm and the “primal-dual” method \cite{bolossom_Zvi1996}. 
Since each step reduces the problem size by roughly half, assuming the maximum degree of the
coupling graph is bounded by a constant, the number of clustering iterations is bounded by $\log n$, 
resulting in an overall complexity for the clustering stage of $O(n^3 \log n +g \log n)$. 


\subsection{Coarsest Level Solving and Refinement}
\label{subsec:refinement}
After obtaining the coarser level circuit and coupling graph, we need to perform refinement to derive a high-quality initial mapping at the coarsest level. 
Additionally, refinement is required after each interpolation step to improve mappings at all intermediate levels. In \cite{lin2024mlqls}, the authors address this by solving the coarsest level optimally using the OLSQ2\cite{lin2023olsq2}, an SMT-based solver, and propose a heuristic algorithm, sRefine, to handle refinement across all intermediate and fine levels. While the SMT solver ensures optimality at the coarsest level, its runtime can still be significant for practical use, even with a fixed problem size. Meanwhile, sRefine, though scalable, is reported to be the primary runtime bottleneck in ML-QLS because of its simulated annealing-based initial mapping and A*-based SWAP insertion, and it could take hours for the program to terminate. Hence, to improve the runtime performance of ML-QLS, a more efficient refinement algorithm is needed. 

Among existing heuristic methods, SABRE emerges as a promising alternative for its effectiveness and efficiency. 
As reported in \cite{qubikos}, SABRE has the smallest optimality gap across different hardware architectures with the shortest runtime, typically terminating in seconds. 
On a device with 18 qubits, it attains an average optimality gap of only 1.09X, making it well-suited for optimizing the mapping at the coarsest level, where the problem size is typically small. By replacing OLSQ2 with SABRE at the coarsest level, we are able to significantly improve runtime performance while still obtaining near-optimal solutions. The design of SABRE also makes it suitable for refining the mapping at each intermediate level. SABRE could take a list of initial mapping and use techniques like bidirectional search, to improve the mapping obtained from interpolation. Therefore, we adopt SABRE as the heuristic for refinement at both the coarsest and intermediate levels. At the coarsest level, we invoke SABRE with randomized initial mappings, aiming to explore a diverse solution space and retain the best-performing result for interpolation. Although the number of trials at this level should be set to a relatively large number, the overall runtime remains efficient due to the small problem size at the coarsest level. At each intermediate level, we provide SABRE with a list of mappings from interpolation to refine. We also include a small number of randomized trials to allow the exploration of other solution spaces since the interpolated mappings may no longer represent globally optimal solutions. However, this random trial is set to be a constant number to avoid exponential growth of the search space. 

\subsection{Interpolation} 
\label{subsec:interpolation}
From the coarsest level, the interpolation algorithm will project the solution back to the finer one, which will then be enhanced via the refinement procedure. 
The interpolation must satisfy two key properties:
First, it must produce a valid mapping, by ensuring each program qubit is assigned to a unique, unoccupied physical qubit. As discussed in Section~\ref{subsec:clustering}, a challenge arises when a program cluster is mapped to a physical cluster with fewer physical qubits than needed, potentially resulting in multiple program qubits being assigned to the same physical qubit, causing an invalid mapping.
Second, the interpolation should preserve the structure of the coarser-level mapping as much as possible. If a program cluster is mapped to a physical cluster, then its constituent program qubits should ideally be mapped to physical qubits within that same physical cluster. This preserves locality and carries over information from the coarser level. However, maintaining structural consistency must not compromise mapping validity. If a one-to-one mapping within the cluster is not possible, the remaining program qubits should be mapped to physical qubits in the neighborhood of the target cluster.

To meet these criteria, we formulate interpolation as a linear assignment problem. The cost of assigning a program qubit to a physical qubit is defined based on its proximity to the coarse-level mapping: zero if the physical qubit belongs to the corresponding physical cluster, and otherwise equal to the shortest distance to any qubit in that cluster. The total cost is the sum of all individual assignment costs. This problem can be solved optimally using the Jonker–Volgenant algorithm\cite{linearAssignment} with time complexity $O(n^3)$. For every coarser-level mapping, there are multiple optimal interpolations, and we will pass a fixed number of them to the refinement algorithm. 

\subsection{Time Complexity Analysis}
\label{subsec:runtime}

Given a circuit with size $g$ and a device coupling graph with size $n$, the time complexity for \mlqls\ is calculated as follows:
First, the initial embedding stage has a worst-case complexity of $O(n^3)$, and it is only invoked once. At the iterative clustering stage, the complexity is $O(n^3 \log n +g \log n)$.
The coarsest level solving uses SABRE, which has a complexity of $O(n^{2.5}g)$~\cite{li_sabre_2019}. The interpolation algorithm has a complexity of $O(n^3)$, 
and SABRE is employed for refinement as well. 
Thus, the combined complexity for iterative interpolation and refinement is $O((gn^{2.5}+n^3)\log n)$. 
Summing all components, the overall complexity for the multilevel framework is $O((gn^{2.5}+n^3)\log n) \approx O(gn^{2.5}\log n)$ assuming $g\gg n$. When we run $d$ cycles, the complexity is $O(dgn^{2.5}\log n)$.

\section{Evaluation}
\label{sec:evaluation}

The proposed algorithm is implemented in Python.
We employ Qiskit (v1.2.4) for LightSabre, NetworkX (v3.1) for maximum-cardinality matching
with maximum weight, 
and Scipy (v1.11.1) for Jonker-Volgenant algorithm. 
Experiments are conducted on an Apple M2 Pro chip with a 12-core CPU. 

We compared ML-SABRE with the state-of-the-art QLS tools, ML-QLS and SABRE.
Our benchmark circuits come from three benchmark sets: QASMBench~\cite{QASMBench}, a known-optimal depth benchmark suite QUEKO~\cite{tan2020queko}, and a known-SWAP-count benchmark suite QUBIKOS~\cite{qubikos}. 
We evaluate our approach on two coupling graphs: a 105-qubit grid architecture based on Google Willow~\cite{google2024willow}, and a 127-qubit heavy-hexagon architecture based on IBM Eagle~\cite{chow2021ibmeagle}.

\subsection{Main Result}
\label{subsec:main_result}
In this experiment, we evaluate \mlqls \space against two other QLS tools: SABRE, and ML-QLS. The circuit used for this experiment has a two-qubit gate count ranging from 31 to 3400.
The configuration for \mlqls \space is as follows: a fixed random seed of 0, 10 multilevel cycles, 500 trials for solving at the coarsest level, 100 interpolations, and one random trial per level. 
SABRE is configured with 2,500 trials and the same random seed. For ML-QLS, we impose a one-hour runtime limit. 

The results for QASMBench circuits on the Eagle architecture are shown in Table~\ref{tab:ibm_eagle_comparison}. 
\mlqls\ consistently outperforms SABRE and ML-QLS both in solution quality and runtime. 
On average, \mlqls\ exhibits a 65\% SWAP count reduction and 17\% depth reduction while attaining a 2.5$\times$ speedup in compilation time compared to SABRE. 
Compared with ML-QLS, \mlqls\ reduces the SWAP count by 45\% and circuit depth by 12\% with a significant speedup up to 240$\times$ in compilation time.
Despite ML-QLS delivering a comparable SWAP count for circuits such as GHZ, Ising, and cat state, its solving time is significantly longer, while  \mlqls\ takes less than one second. 

\mlqls\ does not consistently outperform SABRE and ML-QLS across all instances on the Willow architecture, as shown in Table~\ref{tab:qasmb_willow_results}.
Specifically, SABRE achieves the best performance on the QFT circuit, while ML-QLS yields the best results on the multiplier and DNN circuits.
Nevertheless, \mlqls\ achieves the best performance on average, decreasing the average SWAP count by 45\% and circuit depth by 2\% compared to SABRE  with a 3$\times$ speedup in compilation time.
Compared to ML-QLS, \mlqls\ reduces the SWAP count by 15\% and circuit depth by 2\%, with 315$\times$ speedup in compilation time.
These results highlight the scalability and efficiency of \mlqls, demonstrating its ability to reduce both SWAP count and runtime simultaneously.

The greater improvement of \mlqls\ on IBM Eagle compared to Google Willow can be attributed to differences in hardware topology and clustering strategy.
Since our clustering only groups adjacent physical qubits, coarser-level coupling graphs for the Eagle architecture resemble chains and cycles. 
These structures, though not identical to the heavy-hex layout, still retain good representation of the original coupling graph.
The sparse connectivity of the heavy-hex structure makes the hardware topology the dominant factor in clustering decisions.
This allows refinement steps to effectively exploit the preserved structure to optimize layout.
In contrast, Willow’s dense 2D grid connectivity offers many potential ways to form clusters among neighboring qubits, enabling program qubit interactions to dominate clustering decisions.
However, this flexibility can lead to coarser-level coupling graphs that significantly diverge from the hardware topology.

\begin{table*}[t]
\centering
\begin{tabular}{l|ccc|ccc|ccc}
\toprule
\multirow{2}{*}{Circuit} & \multicolumn{3}{c|}{SABRE (2500 trials)} & \multicolumn{3}{c|}{ML-QLS} & \multicolumn{3}{c}{\mlqls\ (10 cycles)} \\
 & SWAP & Depth & Runtime (s) & SWAP & Depth & Runtime (s) & SWAP & Depth & Runtime (s) \\
\midrule
adder\_n118      & 438 & 1187 & 7.57 & 550 & 1038 & 1277.38 & \textbf{264} & \textbf{913} & \textbf{3.75} \\
bv\_n70          & 28 & 131 & \textbf{0.55} & 44 & 142 & 20.86 & \textbf{27} & \textbf{128} & 1.05 \\
cat\_n65         & 21 & 100 & 0.63 & \textbf{0} & \textbf{68} & 12.06 & \textbf{0} & \textbf{68} & \textbf{0.03} \\
ghz\_n78         & 23 & 117 & 0.67 & \textbf{0} & \textbf{81} & 26.77 & \textbf{0} & \textbf{81} & \textbf{0.03} \\
ising\_n98       & 23 & 22 & 1.56 & \textbf{0} & \textbf{11} & 298.81 & \textbf{0} & \textbf{11} & \textbf{0.04} \\
dnn\_n51         & 127 & 508 & 2.52 & 178 & 413 & 263.59 & \textbf{113} & \textbf{418} & \textbf{2.26} \\
knn\_n67         & 113 & 477 & 1.84 & 168 & 475 & 340.22 & \textbf{87} & \textbf{456} & \textbf{1.55} \\
multiplier\_n45  & 1465 & 3439 & 20.43 & 1914 & 3771 & 727.18 & \textbf{1435} & \textbf{3252} & \textbf{18.34} \\
qft\_n63         & 1803 & 1040 & \textbf{22.61} & 2772 & 1850 & 3600.00 & \textbf{1780} & \textbf{949} & 22.88 \\
qugan\_n71       & 212 & 634 & 3.94 & 236 & 626 & 389.07 & \textbf{165} & \textbf{596} & \textbf{2.73} \\
swap\_test\_n83   & 151 & 595 & 2.36 & 239 & 646 & 364.98 & \textbf{118} & \textbf{563} & \textbf{2.44} \\
wstate\_n118     & 85 & 511 & 3.28 & 290 & 662 & 1026.60 & \textbf{13} & \textbf{486} & \textbf{1.79} \\
\midrule
Geo. Ratio       & 2.87 & 1.21 & 2.54 & 1.83 & 1.14 & 239.80 & \textbf{1.00} & \textbf{1.00} & \textbf{1.00} \\
\bottomrule
\end{tabular}
\caption{QASMBench Evaluation Results on IBM Eagle. The number after the underscore represents the number of qubits in the circuit. For example, adder\_n118 means a quantum adder with 118 qubits.}
\label{tab:ibm_eagle_comparison}
\end{table*}

\begin{table*}[t]
\centering
\begin{tabular}{l|ccc|ccc|ccc}
\toprule
\multirow{2}{*}{Circuit} & \multicolumn{3}{c|}{SABRE (2500 trials)} & \multicolumn{3}{c|}{ML-QLS} & \multicolumn{3}{c}{\mlqls\ (10 cycles)} \\
 & SWAP & Depth & Runtime (s) & SWAP & Depth & Runtime (s) & SWAP & Depth & Runtime (s) \\
\midrule
adder\_n64      & 144 & 542 & 3.23 & \textbf{91} & \textbf{487} & 233.88 & 125 & 604 & \textbf{3.00} \\
bv\_n70         & 20 & 101 & \textbf{0.52} & 34 & 103 & 29.99 & \textbf{11} & \textbf{97} & 0.90 \\
cat\_n65        & 4 & 76 & 0.75 & \textbf{0} & \textbf{68} & 18.01 & \textbf{0} & \textbf{68} & \textbf{0.02} \\
ghz\_n78        & 4 & 89 & 0.49 & \textbf{0} & \textbf{81} & 45.53 & \textbf{0} & \textbf{81} & \textbf{0.05} \\
ising\_n98      & 5 & 16 & 1.62 & \textbf{0} & \textbf{11} & 750.30 & \textbf{0} & \textbf{11} & \textbf{0.05} \\
dnn\_n51        & 76 & 396 & 2.25 & \textbf{62} & \textbf{338} & 1159.86 & 66 & 376 & \textbf{2.18} \\
knn\_n67        & 77 & 420 & 1.67 & 108 & 447 & 146.39 & \textbf{58} & \textbf{383} & \textbf{1.48} \\
multiplier\_n45 & 1071 & 3203 & 18.03 & \textbf{1050} & 3424 & 1291.14 & 1075 & \textbf{3169} & \textbf{12.16} \\
qft\_n63        & \textbf{1275} & \textbf{1302} & \textbf{19.35} & 1483 & 2151 & 1261.85 & 1307 & 1451 & 21.14 \\
qugan\_n71      & 126 & 606 & 3.28 & 126 & \textbf{600} & 629.86 & \textbf{124} & 660 & \textbf{2.46} \\
swap\_test\_n83  & 109 & 531 & 2.20 & 120 & 562 & 200.81 & \textbf{76} & \textbf{487} & \textbf{1.45} \\
wstate\_n76     & 4 & 309 & 1.45 & \textbf{0} & \textbf{305} & 229.94 & \textbf{0} & \textbf{305} & \textbf{0.09} \\
\midrule
Geo. Ratio      & 1.83 & 1.02 & 2.92 & 1.18 & 1.03 & 315.31 & \textbf{1.00} & \textbf{1.00} & \textbf{1.00} \\
\bottomrule
\end{tabular}
\caption{QASMBench evaluation on the Willow architecture.}
\label{tab:qasmb_willow_results}
\end{table*}

\subsection{Optimality Study}
\label{subsec:optimality_study}
This section reports the optimality gap of \mlqls\ and SABRE with respect to both circuit depth and SWAP count.
To have comparable runtime and allow \mlqls\ to explore a broader solution space,
we increase the number of random trials per level to 100,
while using the same configuration in Section~\ref{subsec:main_result} for SABRE. We only compare \mlqls\ against SABRE because in \cite{qubikos}, the author demonstrated that SABRE has the best overall performance compared to other tools, including ML-QLS.

\begin{figure}[ht]
\centering

\begin{subfigure}[t]{\linewidth}
    \centering
    \captionsetup{skip=-2pt}
    \includegraphics[width=\linewidth]{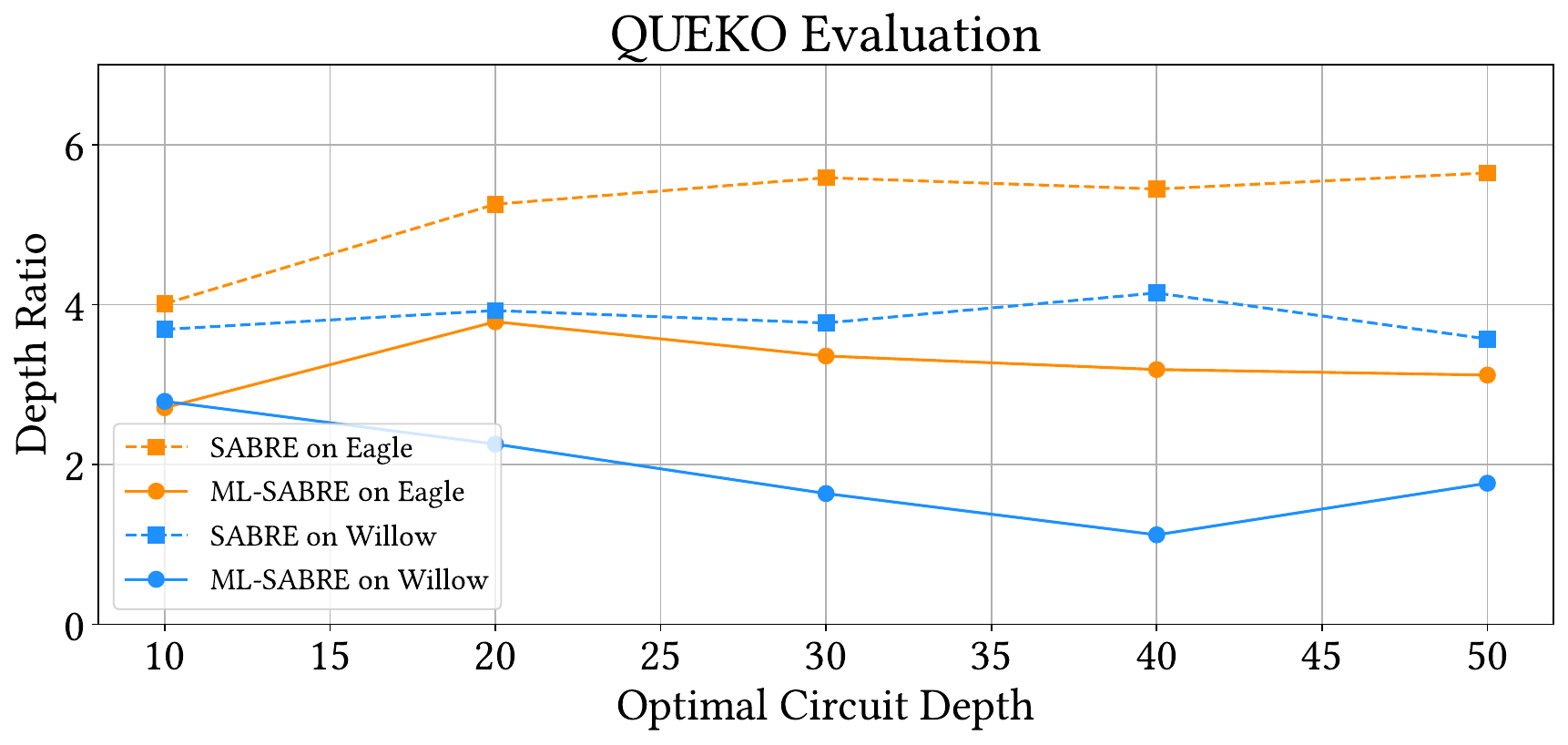}
    \caption{}
    \label{fig:queko}
\end{subfigure}
\hfill
\begin{subfigure}[t]{\linewidth}
    \centering
    \captionsetup{skip=-2pt}
    \includegraphics[width=\linewidth]{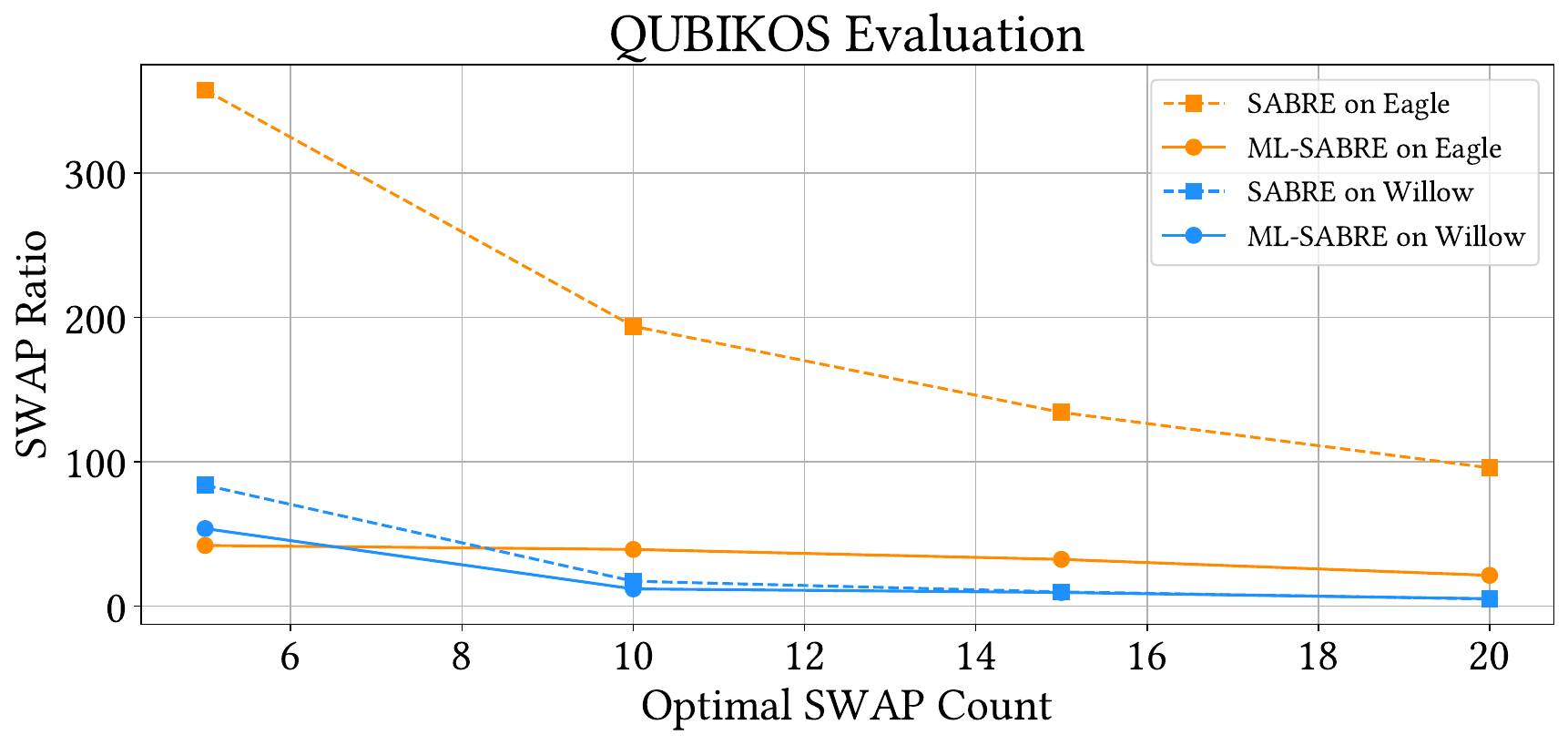}
    \caption{}
    \label{fig:qubikos}
\end{subfigure}

\caption{Optimality study on the Eagle architecture and Willow architecture.
(a) QUEKO.
(b) QUBIKOS.
}
\label{fig:optimality}
\end{figure}

\subsubsection{Circuit Depth}
 \label{queko_evaluation}
We generate QUEKO circuits with optimal depths ranging from 10 to 50, with 10 random circuits per depth. We set the number of qubits to be equal to the device size.  The number of two-qubit varies as the depth change. For circuits with depth 10, 20, 30, 40, and 50, the number of two-qubit gates are 255, 509, 763, 1017, and 1270, respectively. 
The depth ratio is computed as the obtained result divided by the optimal value, and we report the mean ratio over the 10 circuits.
Overall, \mlqls\ realizes a smaller optimality gap than SABRE. 
On the Eagle architecture, SABRE has an average gap of 5.07$\times$, while \mlqls\ reduces it to 3.26$\times$, a 35.8\% improvement. 
On the Willow architecture, the average gaps are 3.88$\times$ for SABRE and 1.95$\times$ for \mlqls, yielding a 49.8\% improvement.







\subsubsection{SWAP Count}
\label{qubikos_evaluation}
We generate QUBIKOS circuits with optimal SWAP count 5, 10, 15, and 20, with 10 circuits for each value. Again, we set the number of qubits in the circuit to be equal to the device size.
The circuit size is limited to 4000 two-qubit gates. 
Similarly, the SWAP ratio is calculated as the obtained SWAP count divided by the optimal value.
The reported ratio is the mean ratio over 10 circuits. 
Figure~\ref{fig:qubikos} exhibits the evaluation results on both architectures.  
\mlqls\ achieves a remarkable improvement of 82.7\% by decreasing the optimality gap from 195.4x to 33.76$\times$ on the Eagle architecture.
On the Willow architecture, the average optimality gaps are 28.96$\times$ for SABRE and 20.07$\times$ for \mlqls, which is a $30.7\%$ reduction.

In summary, the optimality study shows that \mlqls\ closes the optimality gap for both circuit depth and SWAP count. 
Interestingly, \mlqls\ has a greater SWAP count reduction on IBM Eagle, while yielding more significant circuit depth reduction on Google Willow. 
We think this contrast is a result of the differences in the coupling graph connectivity. 
On a densely connected architecture like Willow, there are typically multiple shortest SWAP paths available to bring non-adjacent qubits together. If some of these paths are partially blocked by ongoing circuit operations, alternative paths can often be used to schedule SWAPs in parallel with the rest of the circuit, allowing for a greater depth reduction even without a proportional decrease in SWAP count.
In contrast, the sparse connectivity of IBM Eagle limits the number of viable paths between distant qubits. As a result, SWAP operations are more likely to be serialized, since any interference from existing gate operations can delay SWAP scheduling. This makes it harder to overlap SWAPs with circuit operations, leading to a greater increase in overall circuit depth.
Additionally, Eagle’s limited connectivity results in a larger SWAP count optimality gap, giving \mlqls\ more opportunity for optimization.


\subsection{Ablation Study}
\label{subsec:ablation_study}
We investigate the effectiveness of each setting in \mlqls\ through four experiments: (1) initial embedding, (2) clustering termination criterion, (3) refinement configuration and (4) the number of V-cycles.
Each experiment is repeated with five random seeds (1–5), and results are averaged, except for the V-cycle experiment, which utilizes seed 0. 
For clarity, we use a subset of QASMBench circuits and run all experiments on the IBM Eagle architecture.


\begin{figure}
\centering
\includegraphics[width=1\linewidth]{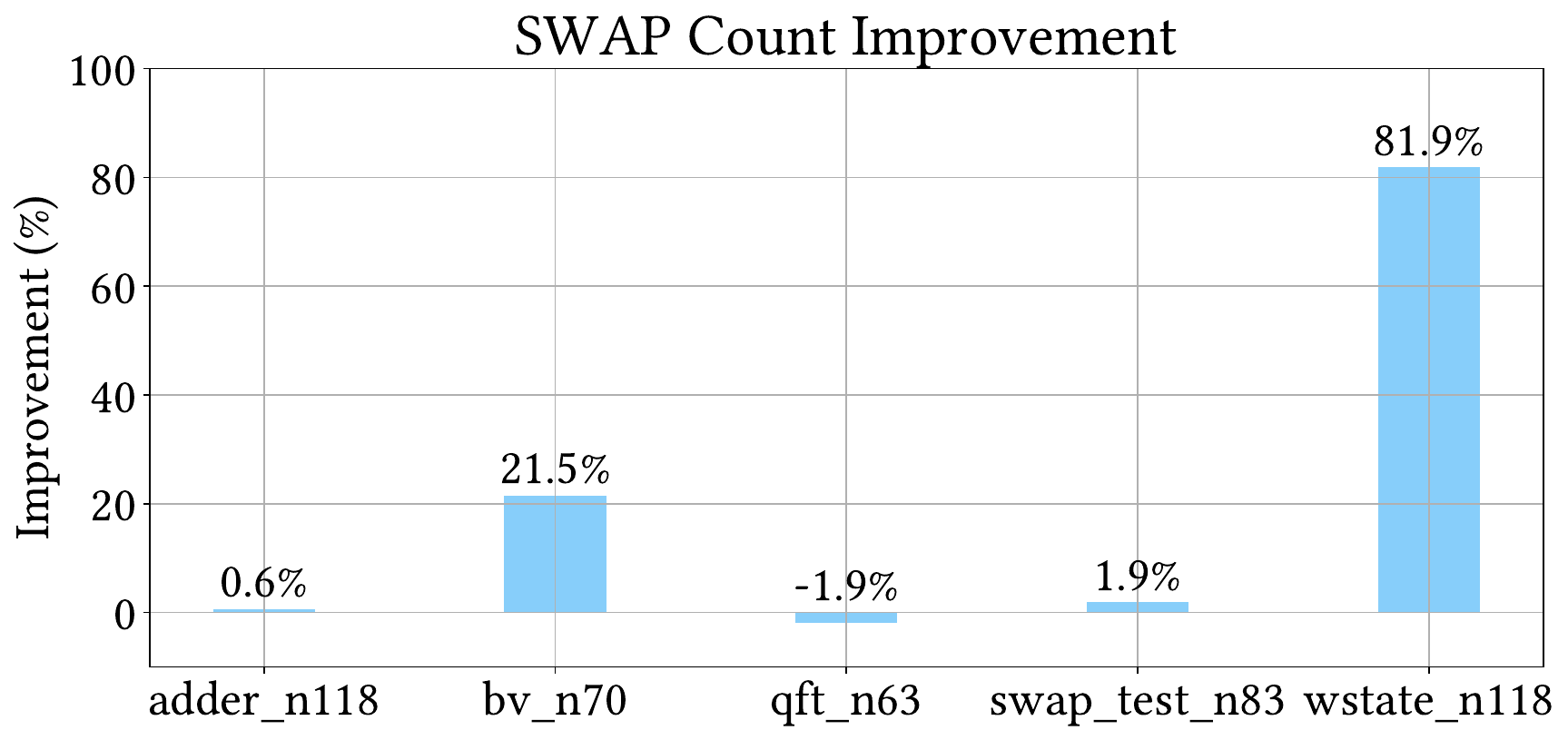}

\caption{SWAP count improvement when employing our initial embedding strategy compared to random mapping. 
}
\label{fig:initialtest}
\end{figure}

\subsubsection{Initial Embedding}
\label{subsec:initial_embedding_exp}
To assess the impact of initial embedding, we compare the method described in Section~\ref{subsec:initial_embedding} against random initial mappings on five representative circuits: adder, QFT, BV, W-state preparation, and SWAP test.
Figure~\ref{fig:initialtest} illustrates the average improvement over the random baseline.
The benefit of initial embedding varies by circuit type: significantly boosting the performance for BV and W-state preparation, performing worse for QFT, and having minimal impact on the adder and SWAP test circuits. 
These results highlight the importance of circuit structure in determining the effectiveness of initial embedding.
Overall, the method provides a favorable starting point for \mlqls.


\begin{figure}
\centering
\includegraphics[width=1\linewidth]{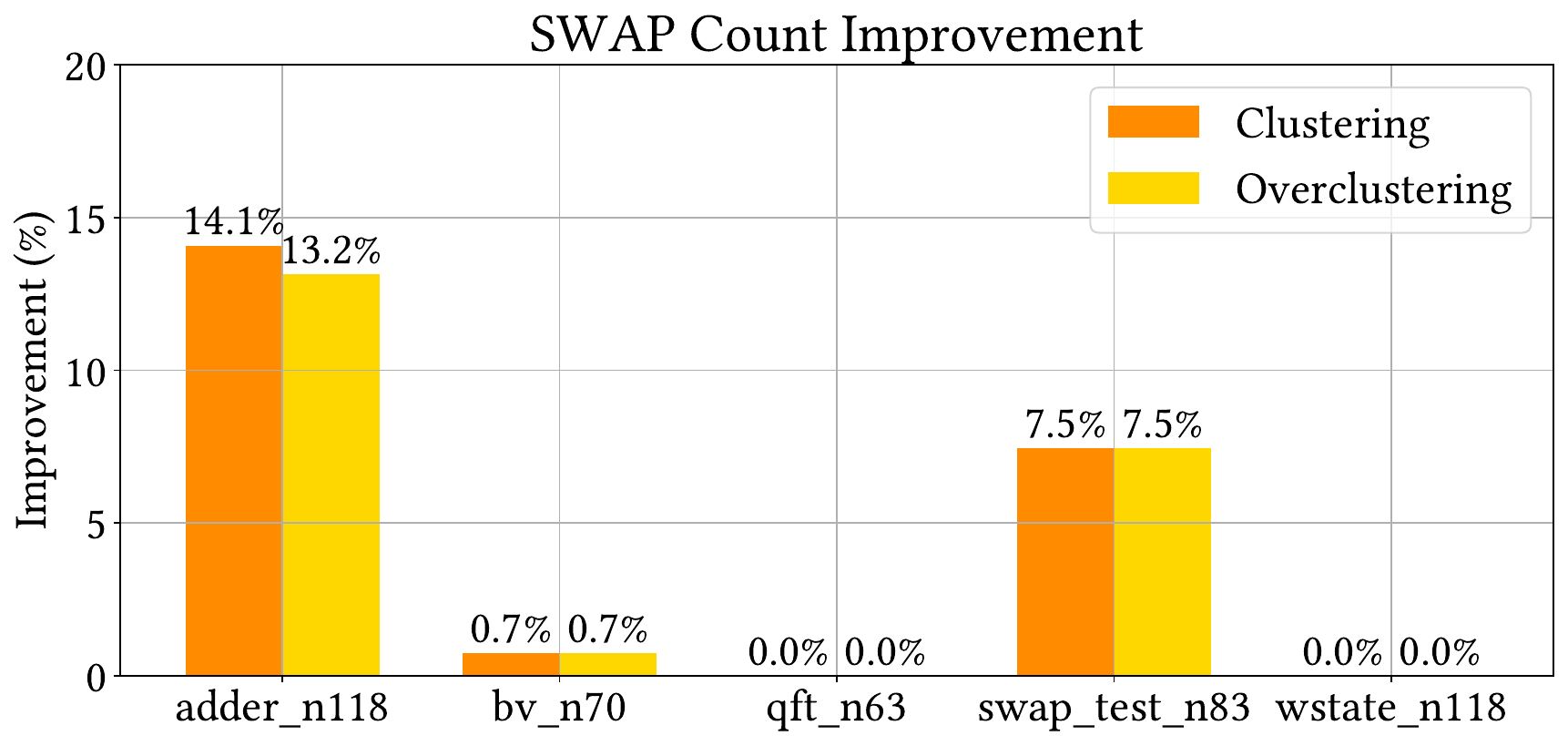}
\caption{SWAP count improvement with different clustering termination criterion. 
The baseline is under-clustering.
The orange bars represent our clustering strategy, and the red bars are for the over-clustering case.}
\label{fig:clustering_test}
\end{figure}

\subsubsection{Clustering Termination Criterion}
\label{subsec:clustering_termination}
As described in Section~\ref{subsec:clustering}, our clustering strategy terminates when it produces a perfect but trivial mapping, and we use the previous level as the coarsest level.
To evaluate this choice, we compare it against two alternatives: over-clustering, which continues until a trivial mapping is reached and used directly, and under-clustering, which performs only one clustering step. 
We utilize under-clustering as the baseline to measure relative improvements.

The result in Figure~\ref{fig:clustering_test} reveal that
our strategy achieves the highest improvement in the Adder circuit. 
For the BV and SWAP test circuits, both our approach and the over-clustering strategy perform equally well. 
In the QFT and W-state circuits, all three strategies yield identical results. This is probably due to the fact that the initial mapping is already good enough, meaning the solver cannot find a better solution at the coarsest level. It does not matter if it is normal or over-clustering.
Overall, the current clustering termination strategy consistently delivers the best average performance across benchmarks.




\begin{figure}[htbp]
\centering

\begin{subfigure}[htbp]{0.48\linewidth}
    \centering
    \includegraphics[width=\linewidth]{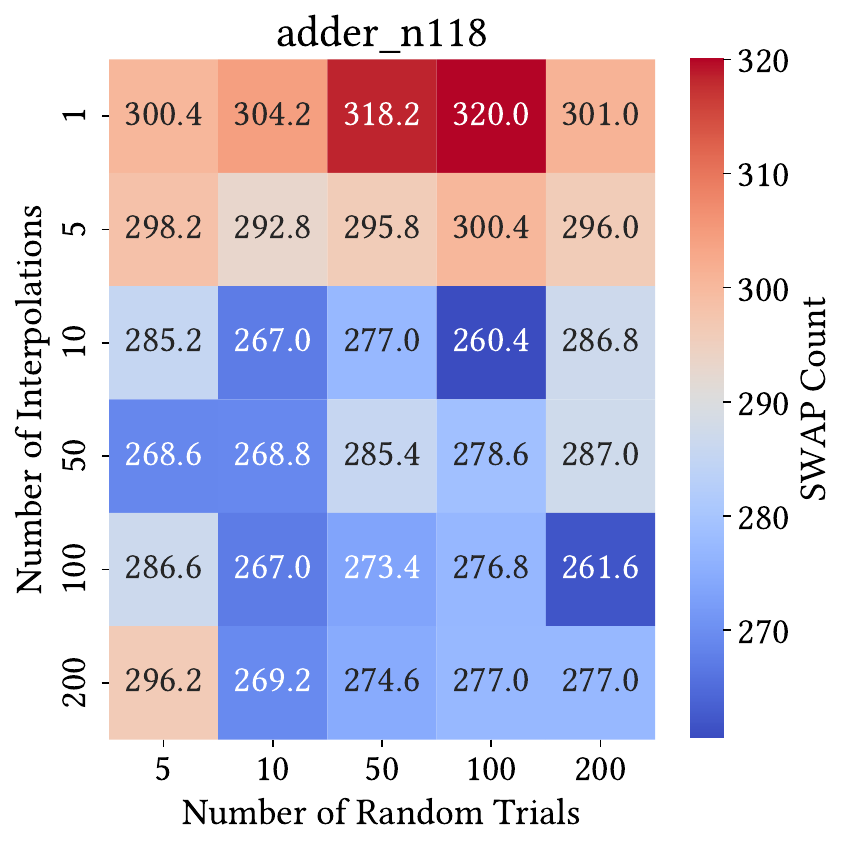}
    \caption{}
    \label{fig:trial_1}
\end{subfigure}
\hfill
\begin{subfigure}[htbp]{0.48\linewidth}
    \centering
    \includegraphics[width=\linewidth]{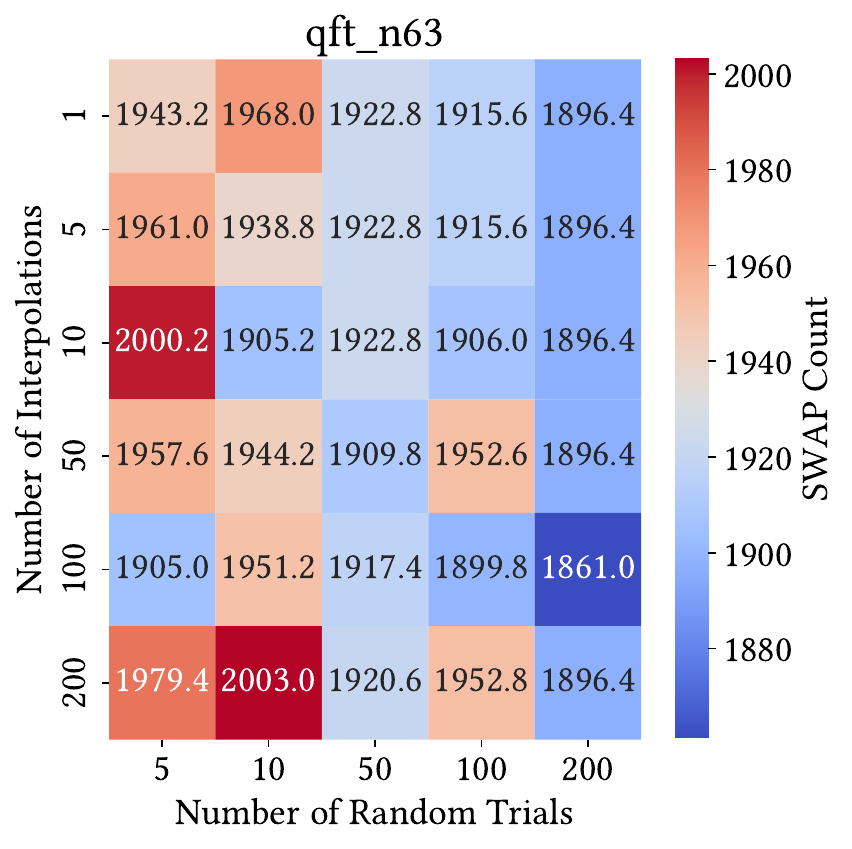}
    \caption{}
    \label{fig:trial_2}
\end{subfigure}
\caption{Different refinement configurations.
(a) Adder circuit with 118 qubits. 
(b) QFT circuit with 63 qubits.}
\label{fig:trial_test}
\end{figure}

\subsubsection{Refinement Configurations}
\label{subsec:trial_number}
At each level, we can enhance the refinement process by increasing either the number of interpolations or random trials. This permits a more thorough exploration of the solution space: interpolations guide the search based on the coarser-level solution, while random trials allow for unconstrained search. 
This experiment evaluates the impact of different refinement configurations by testing  combinations of interpolation and random trial counts, using the values 1, 5, 10, 50, 100, and 200. 
For simplicity, we present the results for the Adder and QFT circuits, visualized as heat maps in Figure~\ref{fig:trial_test}. 
Each grid value indicates the average SWAP count, and the color gradient highlights performance.

\begin{figure}
\centering
\includegraphics[width=\linewidth]{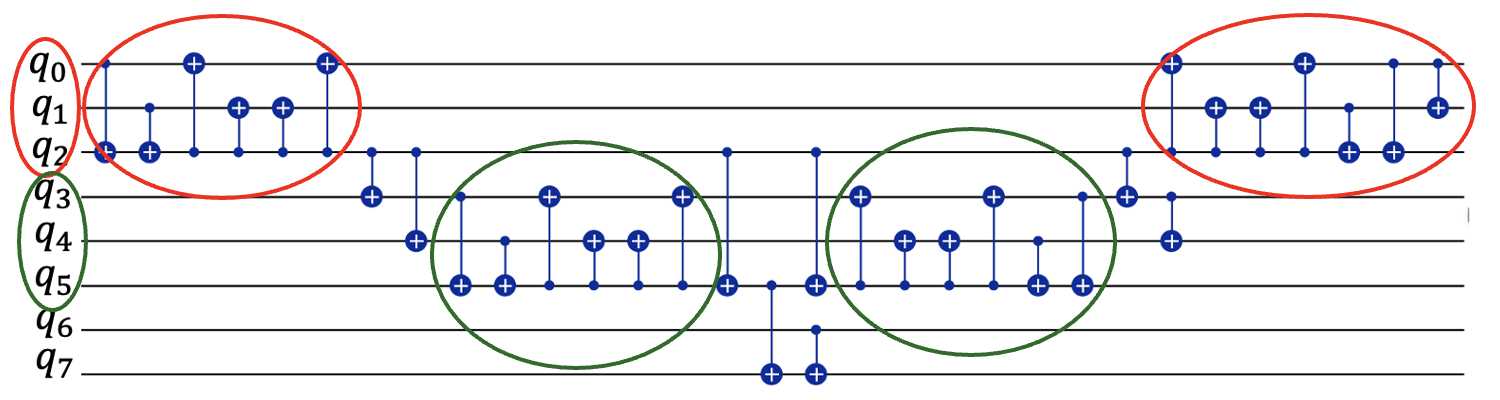}

\caption{An 8-qubit quantum adder circuit showing only two-qubit interactions for clarity. Program qubits grouped in red and green circles exhibit frequent interactions within their respective groups. The subcircuits enclosed by the same color highlight these localized interactions. }
\label{fig:adder}
\end{figure}

In general, we saw a decreasing trend for SWAP count along the diagonal of the heat map, meaning the solution quality will improve as the number of interpolations and random trials increases.
However, the differences between the two figures emphasize how circuit structure affects the effectiveness of refinement strategies. 
For the Adder circuit (Figure~\ref{fig:trial_1}), 
the first two rows show higher SWAP counts, indicating that increasing interpolations is more effective.
This result suggests that good mappings are well-aligned across levels. 
This observation motivated a closer examination of the adder’s circuit structure. Figure~\ref{fig:adder} depicts the two-qubit gates in an 8-qubit adder. 
Notably, the diagram reveals distinct groups of strongly interacting qubits, highlighted by the red (${q_0, q_1, q_2}$) and green (${q_3, q_4, q_5}$) circles. 
Each group contains subcircuits with repeated local interactions, forming clear clusters that \mlqls\ can effectively identify and exploit.
Therefore, circuits with extensive local and repeated interactions are particularly suited for \mlqls. 

In contrast, the QFT circuit exhibits all-to-all connectivity, where each qubit interacts once with every other qubit. 
This lack of locality and repetition weakens the effectiveness of coarse-level guidance, as the coarse and fine solution spaces are less aligned. 
As shown in Figure~\ref{fig:trial_2}, the higher SWAP counts in the first two columns indicate that random trials have a greater impact, making broader exploration at finer levels more effective and limiting the advantage of \mlqls\ for such globally connected circuits.

\subsubsection{V Cycle Count}
\label{subsec:v_count}

\begin{figure}
\centering
\includegraphics[width=\linewidth]{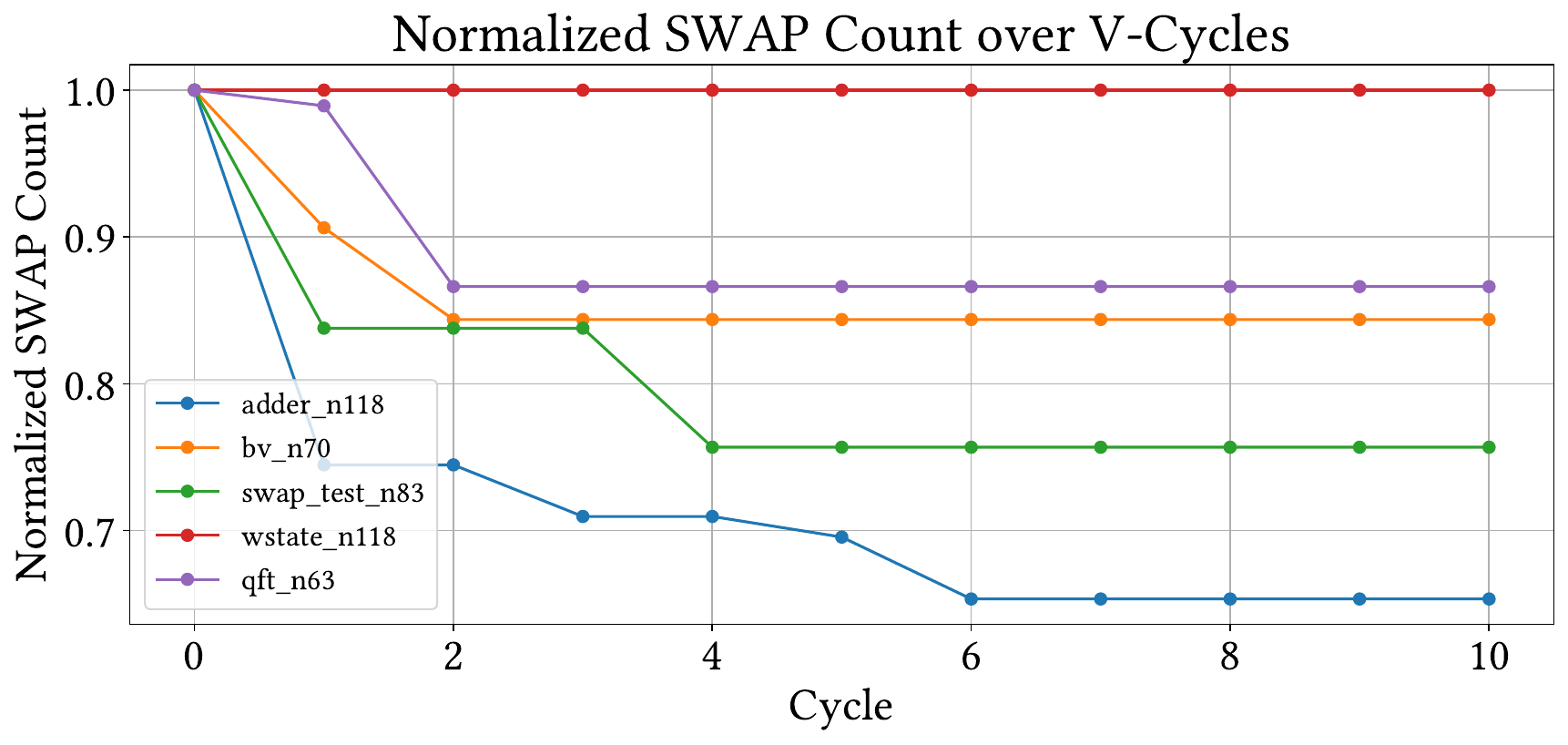}

\caption{Normalized SWAP count versus the number of V cycle. 
SWAP counts are normalized to the baseline at cycle 0, which corresponds to the initial embedding without refinement.}
\label{fig:vcycletest}
\end{figure}

To assess the impact of multiple V cycles, we track the solution obtained for each cycle (Figure~\ref{fig:vcycletest}). 
Cycle 0 corresponds to the SWAP count from the initial embedding via LightSabre, which serves as the baseline. 
Overall, increasing the number of V cycles typically enhances solution quality until it converges to a local minimum.
For instance, the Adder circuit improves up to six cycles before stabilizing, while the SWAP test circuit reaches convergence after just two cycles. 
On the other hand, for circuits like W-state preparation, additional V cycles yield no further gains as the initial embedding is already near-optimal.
Thus, a practical strategy is to monitor convergence and terminate once improvement ceases.

\section{Conclusion}
\label{sec:conclusion}
In this work, we propose \mlqls, a scalable and efficient multilevel framework for QLS that significantly improves both solution quality and runtime. 
Extensive experiments on diverse benchmarks and real quantum architectures demonstrate that \mlqls\ achieves substantial reductions in SWAP count and circuit depth compared to state-of-the-art tools such as SABRE and ML-QLS. 
Our optimality studies further confirm that \mlqls\ consistently narrows the optimality gap for both SWAP count and circuit depth across various device architectures. 
Overall, \mlqls\ advances the state of the art in QLS by delivering high-quality solutions with practical runtime, making it well-suited for future quantum processors with increasing qubit counts and complex connectivity.
\section{Acknowledgment}
This research is supported in part by the National Science Foundation Award No. 2016245 and 2313083.
\bibliographystyle{IEEEtran}
\bibliography{references}

\begin{thebibliography}{10}
\providecommand{\url}[1]{#1}
\csname url@samestyle\endcsname
\providecommand{\newblock}{\relax}
\providecommand{\bibinfo}[2]{#2}
\providecommand{\BIBentrySTDinterwordspacing}{\spaceskip=0pt\relax}
\providecommand{\BIBentryALTinterwordstretchfactor}{4}
\providecommand{\BIBentryALTinterwordspacing}{\spaceskip=\fontdimen2\font plus
\BIBentryALTinterwordstretchfactor\fontdimen3\font minus \fontdimen4\font\relax}
\providecommand{\BIBforeignlanguage}[2]{{%
\expandafter\ifx\csname l@#1\endcsname\relax
\typeout{** WARNING: IEEEtran.bst: No hyphenation pattern has been}%
\typeout{** loaded for the language `#1'. Using the pattern for}%
\typeout{** the default language instead.}%
\else
\language=\csname l@#1\endcsname
\fi
#2}}
\providecommand{\BIBdecl}{\relax}
\BIBdecl

\bibitem{rigetti}
\BIBentryALTinterwordspacing
Rigetti computing. [Online]. Available: \url{https://www.rigetti.com}
\BIBentrySTDinterwordspacing

\bibitem{chow2021ibmeagle}
J.~Chow, O.~Dial, and J.~Gambetta, ``{IBM} quantum breaks the 100-qubit processor barrier,'' \emph{IBM Research Blog}, 2021.

\bibitem{google2024willow}
\BIBentryALTinterwordspacing
{Google}, ``{Meet Willow, our state-of-the-art quantum chip},'' 2024, accessed: 2025-04-12. [Online]. Available: \url{https://blog.google/technology/research/google-willow-quantum-chip/}
\BIBentrySTDinterwordspacing

\bibitem{siraichi_qubitallocation_2018}
M.~Y. Siraichi \emph{et~al.}, ``Qubit allocation.''\hskip 1em plus 0.5em minus 0.4em\relax New York, NY, USA: Association for Computing Machinery, Feb. 2018, pp. 113--125.

\bibitem{wille2014optimal}
R.~Wille, A.~Lye, and R.~Drechsler, ``Optimal {SWAP} gate insertion for nearest neighbor quantum circuits,'' in \emph{2014 19th Asia and South Pacific Design Automation Conference (ASP-DAC)}.\hskip 1em plus 0.5em minus 0.4em\relax IEEE, 2014, pp. 489--494.

\bibitem{bhattacharjee2019muqut}
D.~Bhattacharjee, A.~A. Saki, M.~Alam, A.~Chattopadhyay, and S.~Ghosh, ``{MUQUT}: Multi-constraint quantum circuit mapping on {NISQ} computers,'' in \emph{2019 IEEE/ACM International Conference on Computer-Aided Design}.\hskip 1em plus 0.5em minus 0.4em\relax IEEE, 2019, pp. 1--7.

\bibitem{wille2019mapping}
R.~Wille \emph{et~al.}, ``Mapping quantum circuits to {IBM} {QX} architectures using the minimal number of {SWAP} and {H} operations,'' in \emph{Proceedings of the 56th Annual Design Automation Conference 2019}, 2019, pp. 1--6.

\bibitem{tan2020olsq}
B.~Tan and J.~Cong, ``Optimal layout synthesis for quantum computing,'' in \emph{Proceedings of the 39th {International} {Conference} on {Computer}-{Aided} {Design}}.\hskip 1em plus 0.5em minus 0.4em\relax ACM, Nov. 2020, pp. 1--9.

\bibitem{tan2021olsqga}
------, ``Optimal qubit mapping with simultaneous gate absorption,'' in \emph{2021 IEEE/ACM International Conference on Computer Aided Design}.\hskip 1em plus 0.5em minus 0.4em\relax IEEE, 2021, pp. 1--8.

\bibitem{zhang_time-optimal_2021}
C.~Zhang \emph{et~al.}, ``Time-optimal {Qubit} mapping.''\hskip 1em plus 0.5em minus 0.4em\relax New York, NY, USA: Association for Computing Machinery, Apr. 2021, pp. 360--374.

\bibitem{molavi2022satmap}
A.~Molavi \emph{et~al.}, ``Qubit mapping and routing via {MaxSAT},'' in \emph{2022 55th IEEE/ACM International Symposium on Microarchitecture}.\hskip 1em plus 0.5em minus 0.4em\relax IEEE, 2022, pp. 1078--1091.

\bibitem{nannicini2022optimal}
G.~Nannicini \emph{et~al.}, ``Optimal qubit assignment and routing via integer programming,'' \emph{ACM Transactions on Quantum Computing}, vol.~4, no.~1, pp. 1--31, 2022.

\bibitem{lin2023olsq2}
W.-H. Lin \emph{et~al.}, ``Scalable optimal layout synthesis for {NISQ} quantum processors,'' in \emph{2023 60th ACM/IEEE Design Automation Conference}.\hskip 1em plus 0.5em minus 0.4em\relax IEEE, 2023, pp. 1--6.

\bibitem{ho2018cirq}
A.~Ho and D.~Bacon, ``Announcing {Cirq}: an open source framework for {NISQ} algorithms,'' \emph{Google AI Blog}, vol.~18, 2018.

\bibitem{zulehner2018mapping_to_ibm_qx}
A.~Zulehner \emph{et~al.}, ``An efficient methodology for mapping quantum circuits to the {IBM} {QX} architectures,'' \emph{IEEE Transactions on Computer-Aided Design of Integrated Circuits and Systems}, vol.~38, no.~7, pp. 1226--1236, 2018.

\bibitem{web18-ibm-qiskit}
\BIBentryALTinterwordspacing
{IBM}. (2018) {Qiskit}. [Online]. Available: \url{https://qiskit.org/}
\BIBentrySTDinterwordspacing

\bibitem{zulehner_efficient_2019}
A.~Zulehner \emph{et~al.}, ``An {Efficient} {Methodology} for {Mapping} {Quantum} {Circuits} to the {IBM} {QX} {Architectures},'' \emph{IEEE Transactions on Computer-Aided Design of Integrated Circuits and Systems}, vol.~38, no.~7, pp. 1226--1236, Jul. 2019.

\bibitem{siraichi_qubit_2019}
M.~Y. Siraichi \emph{et~al.}, ``Qubit allocation as a combination of subgraph isomorphism and token swapping,'' \emph{Proceedings of the ACM on Programming Languages}, vol.~3, no. OOPSLA, pp. 120:1--120:29, Oct. 2019.

\bibitem{li_sabre_2019}
G.~Li \emph{et~al.}, ``Tackling the qubit mapping problem for {NISQ}-era quantum devices.''\hskip 1em plus 0.5em minus 0.4em\relax New York, NY, USA: Association for Computing Machinery, Apr. 2019, pp. 1001--1014.

\bibitem{murali_formal_2019}
P.~Murali \emph{et~al.}, ``Formal constraint-based compilation for noisy intermediate-scale quantum systems,'' \emph{Microprocessors \& Microsystems}, vol.~66, no.~C, pp. 102--112, Apr. 2019.

\bibitem{sivarajah_tket_2020}
S.~Sivarajah \emph{et~al.}, ``\BIBforeignlanguage{en}{t$|$ket$\rangle$: a retargetable compiler for {NISQ} devices},'' \emph{\BIBforeignlanguage{en}{Quantum Science and Technology}}, vol.~6, no.~1, p. 014003, Nov. 2020.

\bibitem{kole_improved_2020}
A.~Kole \emph{et~al.}, ``Improved {Mapping} of {Quantum} {Circuits} to {IBM} {QX} {Architectures},'' \emph{IEEE Transactions on Computer-Aided Design of Integrated Circuits and Systems}, vol.~39, no.~10, pp. 2375--2383, Oct. 2020.

\bibitem{liu_notallswaparethesame_2022}
J.~Liu \emph{et~al.}, ``Not all {SWAPs} have the same cost: a case for optimization-aware qubit routing,'' in \emph{2022 {IEEE} {International} {Symposium} on {High}-{Performance} {Computer} {Architecture}}, Apr. 2022, pp. 709--725, {ISSN: 2378-203X}.

\bibitem{wu_robust_2022}
T.-A. Wu \emph{et~al.}, ``A robust quantum layout synthesis algorithm with a qubit mapping checker,'' in \emph{Proceedings of the 41st IEEE/ACM International Conference on Computer-Aided Design}, ser. ICCAD '22.\hskip 1em plus 0.5em minus 0.4em\relax New York, NY, USA: Association for Computing Machinery, 2022.

\bibitem{fan_QLSML_2022}
H.~Fan \emph{et~al.}, ``Optimizing quantum circuit placement via machine learning,'' in \emph{Proceedings of the 59th ACM/IEEE Design Automation Conference}, ser. DAC '22.\hskip 1em plus 0.5em minus 0.4em\relax New York, NY, USA: Association for Computing Machinery, 2022, p. 19–24.

\bibitem{huang2022reinforcement}
C.-Y. Huang \emph{et~al.}, ``Reinforcement learning and dear framework for solving the qubit mapping problem,'' in \emph{Proceedings of the 41st IEEE/ACM International Conference on Computer-Aided Design}, 2022, pp. 1--9.

\bibitem{park2022fsqm}
S.~Park \emph{et~al.}, ``A fast and scalable qubit-mapping method for noisy intermediate-scale quantum computers,'' in \emph{Proceedings of the 59th ACM/IEEE Design Automation Conference}, New York, NY, USA, 2022, p. 13–18.

\bibitem{huang2024ctqr}
C.-Y. Huang and W.-K. Mak, ``{CTQr}: Control and timing-aware qubit routing,'' in \emph{2024 29th Asia and South Pacific Design Automation Conference (ASP-DAC)}.\hskip 1em plus 0.5em minus 0.4em\relax IEEE, 2024, pp. 140--145.

\bibitem{huang2024dear}
------, ``Efficient qubit routing using a dynamically-extract-and-route framework,'' \emph{IEEE Transactions on Computer-Aided Design of Integrated Circuits and Systems}, 2024.

\bibitem{lin2024mlqls}
W.-H. Lin and J.~Cong, ``{ML-QLS}: Multilevel quantum layout synthesis,'' \emph{arXiv preprint arXiv:2405.18371}, 2024.

\bibitem{tan2020queko}
B.~Tan and J.~Cong, ``Optimality study of existing quantum computing layout synthesis tools,'' \emph{IEEE Transactions on Computers}, vol.~70, no.~9, pp. 1363--1373, 2020.

\bibitem{qubikos}
\BIBentryALTinterwordspacing
S.~Ping, W.-H. Lin, D.~B. Tan, and J.~Cong, ``Assessing quantum layout synthesis tools via known optimal-swap cost benchmarks,'' 2025. [Online]. Available: \url{https://arxiv.org/abs/2502.08839}
\BIBentrySTDinterwordspacing

\bibitem{alpert1997multilevel}
C.~J. Alpert \emph{et~al.}, ``Multilevel circuit partitioning,'' in \emph{Proceedings of the 34th annual Design Automation Conference}, 1997, pp. 530--533.

\bibitem{cong2004edge}
J.~Cong and S.~K. Lim, ``Edge separability-based circuit clustering with application to multilevel circuit partitioning,'' \emph{IEEE Transactions on Computer-Aided Design of Integrated Circuits and Systems}, vol.~23, no.~3, pp. 346--357, 2004.

\bibitem{chan_enhanced_2003}
T.~Chan \emph{et~al.}, ``An enhanced multilevel algorithm for circuit placement,'' in \emph{{ICCAD}-2003. {International} {Conference} on {Computer} {Aided} {Design} ({IEEE} {Cat}. {No}.{03CH37486})}, Nov. 2003, pp. 299--306.

\bibitem{chen2005ntuplace}
T.-C. Chen \emph{et~al.}, ``{NTU}place: A ratio partitioning based placement algorithm for large-scale mixed-size designs,'' in \emph{Proceedings of the 2005 International Symposium on Physical Design}, 2005, pp. 236--238.

\bibitem{chan2005multilevel}
T.~Chan, J.~Cong, and K.~Sze, ``Multilevel generalized force-directed method for circuit placement,'' in \emph{Proceedings of the 2005 International Symposium on Physical Design}, 2005, pp. 185--192.

\bibitem{cheng2018replace}
C.-K. Cheng \emph{et~al.}, ``{RePlAce}: Advancing solution quality and routability validation in global placement,'' \emph{IEEE Transactions on Computer-Aided Design of Integrated Circuits and Systems}, vol.~38, no.~9, pp. 1717--1730, 2018.

\bibitem{leong2009replace}
D.~Leong and G.~G. Lemieux, ``{Replace}: An incremental placement algorithm for field programmable gate arrays,'' in \emph{2009 International Conference on Field Programmable Logic and Applications}.\hskip 1em plus 0.5em minus 0.4em\relax IEEE, 2009, pp. 154--161.

\bibitem{chan2000multilevel}
T.~F. Chan \emph{et~al.}, ``Multilevel optimization for large-scale circuit placement,'' in \emph{IEEE/ACM International Conference on Computer Aided Design. ICCAD-2000. IEEE/ACM Digest of Technical Papers (Cat. No. 00CH37140)}.\hskip 1em plus 0.5em minus 0.4em\relax IEEE, 2000, pp. 171--176.

\bibitem{chan2005mpl6}
------, ``{mPL6}: A robust multilevel mixed-size placement engine,'' in \emph{Proceedings of the 2005 International Symposium on Physical Design}, 2005, pp. 227--229.

\bibitem{karypis1997multilevel}
G.~Karypis, R.~Aggarwal, V.~Kumar, and S.~Shekhar, ``Multilevel hypergraph partitioning: Application in {VLSI} domain,'' in \emph{Proceedings of the 34th annual Design Automation Conference}, 1997, pp. 526--529.

\bibitem{cong2005thermal}
J.~Cong and Y.~Zhang, ``Thermal-driven multilevel routing for {3-D} {ICs},'' in \emph{Proceedings of the 2005 Asia and South Pacific Design Automation Conference}, 2005, pp. 121--126.

\bibitem{ou2012non}
H.-C. Ou \emph{et~al.}, ``Non-uniform multilevel analog routing with matching constraints,'' in \emph{Proceedings of the 49th Annual Design Automation Conference}, 2012, pp. 549--554.

\bibitem{lin2002novel}
S.-P. Lin and Y.-W. Chang, ``A novel framework for multilevel routing considering routability and performance,'' in \emph{Proceedings of the 2002 IEEE/ACM International Conference on Computer-Aided Design}, 2002, pp. 44--50.

\bibitem{liu2020cugr}
J.~Liu \emph{et~al.}, ``{CUGR}: Detailed-routability-driven 3{D} global routing with probabilistic resource model,'' in \emph{2020 57th ACM/IEEE Design Automation Conference (DAC)}.\hskip 1em plus 0.5em minus 0.4em\relax IEEE, 2020, pp. 1--6.

\bibitem{zou2024lightsabre}
H.~Zou, M.~Treinish, K.~Hartman, A.~Ivrii, and J.~Lishman, ``{LightSABRE}: A lightweight and enhanced {SABRE} algorithm,'' \emph{arXiv preprint arXiv:2409.08368}, 2024.

\bibitem{VF2}
L.~Cordella, P.~Foggia, C.~Sansone, and M.~Vento, ``A (sub)graph isomorphism algorithm for matching large graphs,'' \emph{IEEE Transactions on Pattern Analysis and Machine Intelligence}, vol.~26, no.~10, pp. 1367--1372, 2004.

\bibitem{bv}
\BIBentryALTinterwordspacing
E.~Bernstein and U.~Vazirani, ``Quantum complexity theory,'' \emph{SIAM Journal on Computing}, vol.~26, no.~5, pp. 1411--1473, 1997. [Online]. Available: \url{https://doi.org/10.1137/S0097539796300921}
\BIBentrySTDinterwordspacing

\bibitem{bolossom_Zvi1996}
\BIBentryALTinterwordspacing
Z.~Galil, ``Efficient algorithms for finding maximum matching in graphs,'' \emph{ACM Comput. Surv.}, vol.~18, no.~1, p. 23–38, Mar. 1986. [Online]. Available: \url{https://doi.org/10.1145/6462.6502}
\BIBentrySTDinterwordspacing

\bibitem{linearAssignment}
D.~F. Crouse, ``{On implementing 2D rectangular assignment algorithms}, year={2016}, volume={52}, number={4}, pages={1679-1696}, keywords={Two dimensional displays;Approximation algorithms;Complexity theory;Minimization;MATLAB;Cost function}, doi={10.1109/TAES.2016.140952},'' \emph{IEEE Transactions on Aerospace and Electronic Systems}.

\bibitem{QASMBench}
\BIBentryALTinterwordspacing
A.~Li, S.~Stein, S.~Krishnamoorthy, and J.~Ang, ``{QASMBench: A Low-Level Quantum Benchmark Suite for NISQ Evaluation and Simulation},'' \emph{ACM Transactions on Quantum Computing}, vol.~4, no.~2, Feb. 2023. [Online]. Available: \url{https://doi.org/10.1145/3550488}
\BIBentrySTDinterwordspacing

\end{thebibliography}

\end{document}